\documentclass[aps,prl,twocolumn,showpacs,floatfix,superscriptaddress,nofootinbib]{revtex4-1}

\usepackage{amssymb,amsmath,amstext}                
\usepackage{graphicx}                                               
\usepackage{epstopdf}                                               
\usepackage{color}                                                     
\usepackage{bm}                                                        
\usepackage{appendix}                                              
\usepackage[utf8]{inputenc}
\usepackage{bbold}
\usepackage{bbm}
\usepackage{ulem}
\usepackage{latexsym}
\usepackage[colorlinks=true,citecolor=blue,linkcolor=magenta]{hyperref}

\def\be{\begin{equation}}
\def\ee{\end{equation}}
\def\bea{\begin{eqnarray}}
\def\eea{\end{eqnarray}}
\def\ra{\rangle}

\def\bi{\begin{itemize}}
\def\ei{\end{itemize}}
\def\rmd{\mathrm{d}}

\definecolor{dgreen} {RGB}{78,138,21}

\begin{document}

\title{Many-body localization caused by  temporal disorder}

\author{Marcin Mierzejewski} 
\affiliation{Department of Theoretical Physics, Faculty of Fundamental Problems of Technology, Wroc\l aw University of Science and Technology, 50-370 Wroc\l aw, Poland}
\author{Krzysztof Giergiel} 
\affiliation{
Instytut Fizyki imienia Mariana Smoluchowskiego, 
Uniwersytet Jagiello\'nski, ulica Profesora Stanis\l{}awa \L{}ojasiewicza 11, PL-30-348 Krak\'ow, Poland}
\author{Krzysztof Sacha} 
\affiliation{
Instytut Fizyki imienia Mariana Smoluchowskiego, 
Uniwersytet Jagiello\'nski, ulica Profesora Stanis\l{}awa \L{}ojasiewicza 11, PL-30-348 Krak\'ow, Poland}
\affiliation{Mark Kac Complex Systems Research Center, Uniwersytet Jagiello\'nski, ulica Profesora Stanis\l{}awa \L{}ojasiewicza 11, PL-30-348 Krak\'ow, Poland
}

\date{\today}

\begin{abstract}
The many--body localization  (MBL)  is commonly related to a strong spatial disorder.   
We show that  MBL  may alternatively be generated by adding a  temporal disorder to periodically driven many-body systems.
We reach this conclusion by mapping the evolution of such systems on the dynamics of the time-independent, disordered, Hubbard--like models.
Our result opens the way to experimental studies of MBL in {systems that reveal crystalline structures in the time domain.} In particular,  we discuss two relevant setups which can be implemented in experiments on ultra-cold atomic gases. 
\end{abstract}
\date{\today}

\maketitle

Many-body localization (MBL) \cite{Basko06,Oganesyan07,Znidaric08} appears as one of the most challenging phenomena in the  many-body physics, as manifested by hundreds of papers currently appearing each year in this field (for recent reviews see, e.g., \cite{Huse14,Rahul15}).   Due to numerous theoretical studies which have been carried out in the last decade, it is now possible to identify main hallmarks of the MBL:  {vanishing of dc transport   \cite{berkelbach10,barisic10,agarwal15,lev15,steinigeweg15,barisic16, kozarzewski16};}  absence of thermalisation 
\cite{pal10,serbyn2013,lev14,schreiber15,serbyn15,khemani15,luitz16,gornyi05,altman15,deluca13,gramsch12,deluca14,huse13,Rahul15,rademaker16,chandran15,ros15,eisert15,serbyn141,Rahul15,zakrzewski16}
  accompanied by extremely slow dynamics of  various correlation-functions \cite{monthus10,pal10,luitz16, mierzejewski16, prelovsek16,mondaini15} and the logarithmic growth of the entanglement entropy \cite{Znidaric08,bardarson12,kjall14,serbyn15,luitz16}.  
  
   In contrast to a vast amount of theoretical results, there are only a few experimental studies on the MBL, focused on the suppression of the particle  transport in the cold--atoms \cite{schreiber15,choi16,bordia16,bordia2017_1} or  trapped ions \cite{smith2016}. It  is unexplored whether MBL may be implemented in the solid--state devices where coupling  to other degrees of freedom (e.g. phonons or magnons) may disrupt the localization \cite{Basko06,MOTT1968_1,mondaini15,prelovsek216,parameswaran2017,bonca2017}.
  Consequently, it seems important to find other experimental  setups which host the MBL. 
   
     The main properties of MBL can be {explained via} the presence of  quasi-local integrals of motion \cite{Serbyn13,Huse14,Rahul15, imbrie2017} which  prevent thermalization (in the sense of the eigenvector thermalisation hypothesis \cite{srednicki94}) in a large isolated system.   Thus, MBL stabilizes the dynamics and it is claimed to prevent a driven system from heating\cite{kozarzewski16,abanin2017,abanin2015,Lazarides15,Ponte2015,Kuwahara2016,Bordia2017,Rehn2016}.   This idea was used {in recent experiments} to stabilize the so called discrete (or Floquet) time crystals \cite{Choi2017,Zhang2017} resulting from spontaneous breaking of discrete time translation symmetry in periodically driven systems
\cite{Sacha2015,Khemani16,Keyserlingk16,ElseFTC,Yao2017}. However, MBL is not a necessary condition to observe Floquet time crystals \cite{Sacha2015,Nayak2017,Else17prx,Huang2017}.


In the existing experimental studies,  MBL is caused by strong {\em spatial} disordered \cite{schreiber15,choi16,bordia16,bordia2017_1,smith2016,Choi2017,Zhang2017}.  
In this letter, we {provide a new} perspective and show that in {systems that reveal crystalline structures in time,} MBL can be caused  by {\em  temporal} disorder. 
This is important not only as a matter of principle but also as a guideline to build a new class of systems which may host the MBL.
{The crystalline structure in time means that when we fix position in configuration space,  {then} probability for detection a {particle} at this position reveals periodic crystalline behavior versus time.}
It was already suggested that (single--particle) Anderson localization and many-body superfluid-Mott insulator transition can be studied in the time domain \cite{Sacha15a}.
The aim of this {work  is to demonstrate} that MBL itself can be investigated in time. We show that periodically driven many-body systems in the presence
of a temporal disorder exhibit in the time domain the same localization properties as disordered systems in configuration space, and may thus be many-body localized.
This is obtained by mapping  the relevant Floquet eigenstates onto the eigenstates of a tight-binding model \cite{Guo2013,Sacha15a}. We consider two possible experimental realizations: cold atoms bouncing on an oscillating mirror \cite{Steane95,Buchleitner2002}
and cold atoms distributed along a ring shaped optical trap \cite{Clifford1998,Wright2000,Wang2009}. 

We start {with} a one-dimensional (1D) system containing a single particle of unit mass described by an integrable Hamiltonian $H_0(x,p),$ driven by a periodically changing perturbation $H_1(t)=\lambda g(x)\cos\omega t$ where $\lambda$ and $\omega$ are, respectively, the driving amplitude and frequency. 
{We will derive an effective Hamiltonian of the system within the classical secular approximation \cite{Lichtenberg1992} and then quantize it because it is easier to explain emergence of a crystalline structure in time. However, the same results can be obtained within a fully quantum approach \cite{Berman1977}.}
In {the} classical description, it is convenient to perform a canonical transformation from Cartesian coordinates $(x,p)$ to the action-angle variables $I$ and $\theta$ of the unperturbed system \cite{Lichtenberg1992}. Then, $H_0=H_0(I)$ and the unperturbed {motion is described by} $I=\mathrm{constant} $ and $\theta=\omega_0t+\theta_0$, where
the motion frequency is given by $\omega_0=\frac{\rmd H_0(I)}{\rmd I}.$ When the system is resonantly driven, i.e. the frequency $\omega$ of the external driving fulfills the $s:1$ resonance condition $\omega=s\omega_0$ with {integer} $s$, then, in the rotating frame, $\Theta=\theta-\omega t/s$, the position and the conjugate momentum $P=I-I_s$ are slowly varying variables in the vicinity of the resonant orbit $P\approx 0$. Averaging the Hamiltonian over the {\it fast} time oscillations yields
\be
H\approx H_{\rm sec}= \frac{P^2}{2m}+\lambda g_s(I_s)\cos(s\Theta),
\label{secH_cl}
\ee 
where $m=\left(\frac{\rmd^2 H_0(I_s)}{\rmd I_s^2}\right)^{-1}$ is the effective mass and $g_s(I)$ is the Fourier component of $g(x)=g(\theta,I)=\sum_n g_n(I)e^{in\theta}$ \cite{Lichtenberg1992}.
The classical secular Hamiltonian~(\ref{secH_cl}) is spatially periodic and for $s\gg 1$ it {resembles a} Hamiltonian {of an electron} in a crystal with periodic boundary conditions \cite{Guo2013,Sacha15a}. We turn to its quantum version 
{and consider only} the lowest energy band (that forms when $s$ is big), i.e. we {study} the Hilbert subspace spanned by the Wannier states $w_j(\Theta)$ of the periodic potential in (\ref{secH_cl}).
Then, the wavefunction can be expanded as $\psi=\sum_{j=1}^s{a_jw_j}$ (with $a_j$ arbitrary complex numbers) and the energy of the system is given:
\be
E=\int_{0}^{2\pi}d\Theta \;\psi^*H_{\rm sec}\psi\approx -\frac{J}{2}\sum_{j=1}^s(a_{j+1}^*a_j+c.c.),
\label{tight_cl}
\ee
where $J=-2\int d\Theta w_{j+1}^*H_{\rm sec}w_j$ is the tunneling amplitude of the particle between neighboring potential wells.

We have thus reduced the description of a resonantly driven single particle system to a tight-binding model (\ref{tight_cl})
in the rotating frame \cite{Guo2013,Sacha15a}. When we return to the laboratory frame, a single Wannier state $w_j(\theta-\omega t/s)$ is a localized wavepacket moving along the $s:1$ resonant orbit with a period  $s$ times longer than the driving period $2\pi/\omega$.
It is however not stable in the long time limit as 
it will tunnel (over a time scale $\hbar/J$) to other Wannier states and thus loose its localization properties.  
An eigenstate of the system in the rotating frame corresponds to a Floquet eigenstate of the original periodically driven system. Because the tight-binding Hamiltonian (\ref{tight_cl}) is invariant {under} translation, its
eigenstates are Bloch waves of the type  $\psi_k=\sum_{j=1}^s{\mathrm{e}^{2i\pi jk/s}w_j}$ with $k$ an integer in the range $[0,s\!-\!1]$; the associated energy is $-2J\cos(2\pi k/s).$ In the laboratory frame, this is a train of localized wave-packets 
 $\sum_{j=1}^s \mathrm{e}^{2i\pi jk/s} w_j(\theta-\omega t/s)$. Thus, if we locate a detector in the laboratory frame close to the resonant trajectory, we will observe that the clicking probability changes periodically in time, i.e. the probability becomes significant when each localized wave-packet $w_j(\theta-\omega t/s)$ arrives close to the detector. It shows that our system reveals periodic crystalline structure in the time domain \cite{Sacha15a} similarly  {to} a particle in the presence of a time-independent space periodic potential. Note, that in general, such a periodic behavior is not observed in space, i.e. versus $x$ for a fixed time $t$. Indeed, while the relation between $\Theta$ and $t$ is linear and periodic behavior in $\Theta$ implies periodic behavior in time, the same is not true for $x$ because the canonical transformation between $(x,p)$ and $(\Theta,P)$ is in general non-linear \cite{sacha16}. 

The existence of localized Wannier states evolving 
with a period exactly equal to $s$ times the driving period -- before quantum tunneling sets in -- is robust versus any microscopic imperfection. Indeed, it is based on the classical $s:1$ resonance between the internal frequency and the external driving frequency. The existence of a resonance island with a finite area in the phase space ensures a robust locking of the dynamics {to} the external frequency \cite{Buchleitner2002}. 
The single particle crystalline structure (energy bands and Bloch eigenstates) opens a possibility for realization of many-body crystalline behavior in time where 
the quantum object is not destroyed by a measurement
and the crystal character is preserved for arbitrarily long times, in the thermodynamic limit. 

Both these requirements are met if one considers a many-body system with identical spinless Bosons or spin--$\frac{1}{2}$ fermions  {described by} the same Hamiltonian. One then has to replace the  wave-function $\psi$  in Eq. (\ref{tight_cl}) respectively by a bosonic or fermionic field operators. For simplicity we discuss only the  case of spinless bosons, leaving aside an obvious extension to the case of spinful fermions. If we restrict to the Hilbert subspace spanned by Fock states $|n_1,\dots,n_s\ra$ where $n_j$ is occupation of a Wannier state $w_j$, the many-body system in the rotating frame is described by the Bose-Hubbard Hamiltonian,
\bea
\hat H_0&=&-\frac{J}{2}\sum_{j=1}^s(\hat a_{j+1}^\dagger \hat a_j+h.c.)
+\frac12 \sum_{i,j=1}^s U_{ij}\hat a_i^\dagger\hat a_i\hat a_j^\dagger\hat a_j,
\label{tight_q}
\eea
where 
$U_{ij}=g_0\frac{\omega}{2s\pi}\int_0^{2s\pi/\omega} dt\int_{-\infty}^{\infty}dx |w_i(x,t)|^2|w_j(x,t)|^2$   {is an} effective coupling resulting from contact interactions between ultra-cold atoms with strength $g_0$ that depends on s-wave scattering length and on a transverse confinement of a 3D system \cite{Sacha15a,Guo2016a}. The first part of the Hamiltonian (\ref{tight_q}) is a many-body counterpart of (\ref{tight_cl}) where amplitudes $a_j$ and $a_j^*$ are replaced by annihilation and creation operators $\hat a_j$ and $\hat a_j^\dagger$, respectively. 
In the case of ultra-cold atoms that are the mixture of different kinds of fermions, the many-body Hubbard Hamiltonian looks similar to (\ref{tight_q}) but interactions occur between different species only. 

{Further on, we} focus on the driven many-body system in the presence of disorder.
 {The simplest method of introducing the disorder  is to work in the "time" space and to add }
 a weak perturbation $H'(t)=g(x)f(t)$ where $f(t)$ is time-periodic with the long period $s$ times $2\pi/\omega$ but having random fluctuations during each period, between $t=0$ and $2s\pi/\omega.$ We can expand $H'(t)$ in a Fourier series: $H'(t)=g(x)\sum_{q\ne 0}f_qe^{iq\omega t/s}$ where $f_q=-f_{-q}^*$ are independent random variables. {The many--body Hamiltonian in the rotating--frame  acquires an additional term} 
\begin{equation}
\hat H= \hat H_0 + \sum_{j=1}^s\epsilon_j\hat a_j^\dagger\hat a_j, 
\label{htot}
\end{equation}
 where   $ \hat H_0$ is given by (\ref{tight_q}). Here, $\epsilon_j=\int d\Theta w_j^*V_{\rm dis}(\Theta)w_j$ and $V_{\rm dis}(\Theta)=\sum_{q\ne 0}g_q(I_s)f_{-q}e^{iq\Theta}$ is an effective disordered potential whose statistical properties can be engineered by a choice of a distribution  for random variables $f_q$. In the presence of the perturbation, 
the translational invariance (equivalence of the various $j$ sites) is broken.
 {While the system still possesses Floquet eigenstates, they 
 have $s$ times longer period, $2s\pi/\omega$ }. 
 {The Hamiltonian (\ref{htot}) is valid provided the interaction energy $NU_{ij}$ ($N$ is a total number of bosons) and the disorder $\epsilon_j$ are much smaller than the energy gap between the first and second energy bands of (\ref{secH_cl}). This condition can be easily fulfilled because we {consider}  perturbations of the order of a few tunnelling amplitudes $J$ which is a tiny energy scale.}

The disordered Hubbard model (\ref{htot}) has been the starting point for  majority of experimental studies on MBL.  In particular, such system has been studied in
 \cite{schreiber15,choi16,bordia16,bordia2017_1}  and  \cite{choi16} for spinful fermions and bosons,  respectively. {Theoretical  studies of the disordered fermionic Hubbard model  can  be found, e.g., in Refs. \cite{mondaini15} and \cite{prelovsek216}.}
  The strongest experimental support for  MBL in this model comes from the direct observation of  the density of particles in the real space (e.g. see results for  imbalance in \cite{schreiber15}). The general idea is that the local operator $\hat {\cal I}=\sum_j \alpha_j  \hat a_j^\dagger\hat a_j $ avoids thermalization in that its expectation
value depends on the initial state within the entire experimentally accessible time scale.  
The choice of  coefficients $\alpha_j$ (uncorrelated with $\epsilon_j$) reflects the details of experimental setup and, e.g. the preparation of the initial state. However, the
same quantity can be measured also in {systems that reveal crystalline structure in time} (\ref{htot}) by a detector located close to the resonant trajectory. Below we  discuss in more details two possible experimental implementations of this general idea. {We would like to stress that MBL in driven systems considered here can be observed only in a Hilbert subspace spanned by Wannier modes $w_j$, i.e., in the subspace where  systems are described by the Hamiltonian (\ref{htot}).}

\begin{figure}[h] 	            
\includegraphics[width=0.8\columnwidth]{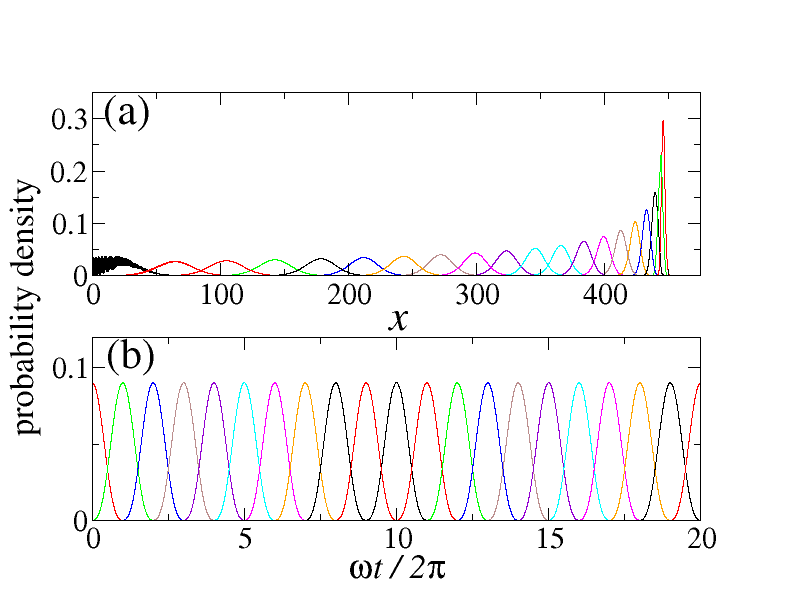}       
\caption{(color online) (a): Wannier states $w_j(x,t)$ that form the basis modes for the tight-binding Hamiltonian (\ref{htot}) in the case of the $20:1$ resonance for $\lambda=0.0122$ and $\omega=2.1$ and for $t=\pi/2\omega$. The gravitational units are used, i.e. $l_0=(\hbar^2/m^2\tilde g)^{1/3}$, $t_0=(\hbar/m\tilde g^2)^{1/3}$ and $E_0=m\tilde g l_0$ for length, time and energy, respectively. Each Wannier state evolves with the period $40\pi/\omega$. Superpositions of the Wannier states form 20 Floquet eigenstates of a single particle bouncing on the oscillating mirror that evolve with the period $2\pi/\omega$. (b): Wannier states as a function of time for a fixed position in the configuration space ($x=448$). This panel illustrates crystalline structure in the time domain. The results have been obtained within the quantum secular approximation \cite{Berman1977}.}
\label{bouncer}   
\end{figure} 

Let us first consider ultra-cold atoms bouncing on a moving mirror in the presence of a gravitational field \cite{Szriftgiser:Mirror:PRL96}. We assume strong transverse confinement so that the description of the system can be reduced to a 1D model. At the beginning, let us describe the single-particle problem. In the non-inertial frame where the mirror is fixed at $x=0,$ an atom moves in the half-space $x\ge 0$ in a time-dependent gravitational field. The static part of the single-particle Hamiltonian is simply $H_0=p^2/2m+m\tilde gx,$ with $\tilde g$ the constant gravitational field.
A mirror oscillating at frequency $\omega$ adds a term $H_1=\lambda x\cos\omega t$  {to} the Hamiltonian. The resonance condition 
$\omega=s\omega_0$ selects motion of an atom with an unperturbed period $2s\pi/\omega$ and an atom is bouncing on the mirror with a vertical
amplitude $h=\tilde g s^2\pi^2/2\omega^2.$ For sufficiently large $\lambda,$ there are $s$ possible bouncing quantum wavepackets \cite{Buchleitner2002} that constitute the Wannier states discussed previously, see Fig.~\ref{bouncer}.
One can easily {introduce} a temporal disorder in the oscillations of the moving mirror, creating an additional disordered Hamiltonian $H'(t)=xf(t)=x\sum_{q\ne 0}f_qe^{iq\omega t/s}.$ 
In the many-body  {system, i.e., for the}  ultra-cold atomic cloud bouncing on the mirror, the atom-atom interaction is responsible for the $U_{ij}$ terms, as in (\ref{tight_q}), that allow us to study many-body transport and localization in this system. While the diagonal terms $U_{ii}$  are {the strongest interactions}, there are {also off-diagonal} contributions arising from the crossing in configuration space of the $i^\mathrm{th}$ wavepacket moving upwards with the $j^\mathrm{th}$ wavepacket moving downwards. These  {off-diagonal} terms are typically an order of magnitude smaller than diagonal ones. For example for the parameters chosen in Fig.~\ref{bouncer}, we get $U_{ii}/g_0=5.92\cdot 10^{-3}$, $U_{ii+1}/g_0=6.9\cdot 10^{-4}$ and $U_{ii+10}/g_0=1.2\cdot 10^{-4}$ while $J=1.9\cdot 10^{-4}$. Slight changes of the amplitude $\lambda$ leave $U_{ij}$ practically intact but significantly change the value of the tunneling amplitude $J$. Alternatively one can change $g_0$ by modification of a transverse confinement of atoms or by changing s-wave scattering length with the help of a Feshbach resonance. Thus, there is an easy way to control the ratios $U_{ij}/J$ in a laboratory.   

{It is believed that MBL is constrained to systems with short--range interactions \cite{yao2014}. However, recent non--perturbative studies  in  \cite{nand2017} indicate that  the long--range interaction alone does not exclude the MBL.  Generally,  the existence of MBL in  systems with non-local interactions is an important but unexplored
 problem. The Hamiltonian (\ref{htot}) is different from that in \cite{nand2017}, however it contains at least a weak non-local component of the many-body interaction.
  These long--range terms are  weak, $U_{i \ne j} \ll U_{ii}$,  hence they should not destroy MBL at least not within a reasonable time scale.}
 
The experiment could be done as follows: firstly, launch a many-body wavepacket from a certain altitude
above a mirror and adjust the vibration frequency of the mirror to match the $1:1$ resonance with the natural bouncing frequency. This creates an atomic wavepacket locked on the external vibration frequency. Secondly, multiply abruptly the
vibration frequency by a factor $s$ and add a "disordered" temporal modulation of the mirror position. It results in an initial many-body state where all atoms occupy one Wannier state. Then, one can monitor
the atomic density at a given position vs. time and observe whether the moving wavepacket remains localized 
or  {it} is transferred to other Wannier wavepackets oscillating at the same frequency but shifted in time, i.e. one can monitor average value of the local operator $\hat {\cal I}$. 

The phase diagram with MBL boundaries has been obtained only for the disordered  Heisenberg model   (see, e.g. \cite{luitz15}) and equivalent
model of spinless fermions.  In the case of the Hubbard model such information is still missing. Moreover, it is by far not obvious whether/when the strict MBL may be observed in the latter system \cite{mondaini15,prelovsek216,bonca2017}.  Therefore,  in order to locate the relevant model parameters one may follow the experimental results for the disorder fermionic Hubbard mode,
e.g. see Fig 4. in \cite{schreiber15}.   Typically,  the many-body interaction $U$ is comparable with $J$, while the disorder strength should be the largest
energy scale in the system. 
As mentioned before, the ratios $U_{ij}/J$ can be easily controlled in a laboratory. Suitable choice of random components $f_q$ in $H'(t)$ allows one to engineer statistical properties of on-site energies $\epsilon_j$ in (\ref{htot}). For example if $f_q=V_0\omega^2q^2s^{-2}e^{i\varphi_q}$ for $|q|\le s$ and zero otherwise, where $\varphi_q=-\varphi_{-q}$ are random numbers chosen uniformly in $[0,2\pi)$, we obtain $V_{\rm dis}(\Theta)=V_0\sum_{q=-s}^se^{i(q\Theta+\varphi_q)}$. Then, employing the central limit theorem, one can show that $\epsilon_j$ are random numbers corresponding to the normal distribution with zero mean and standard deviation $\sigma=\sqrt{2s}V_0$. With the help of the parameter $V_0$ one can control the strength of the disorder in a wide range including $\sigma\gg J$ and investigate the entire phase diagram of the system.
   
\begin{figure}[h] 	            
\includegraphics[width=0.4\columnwidth]{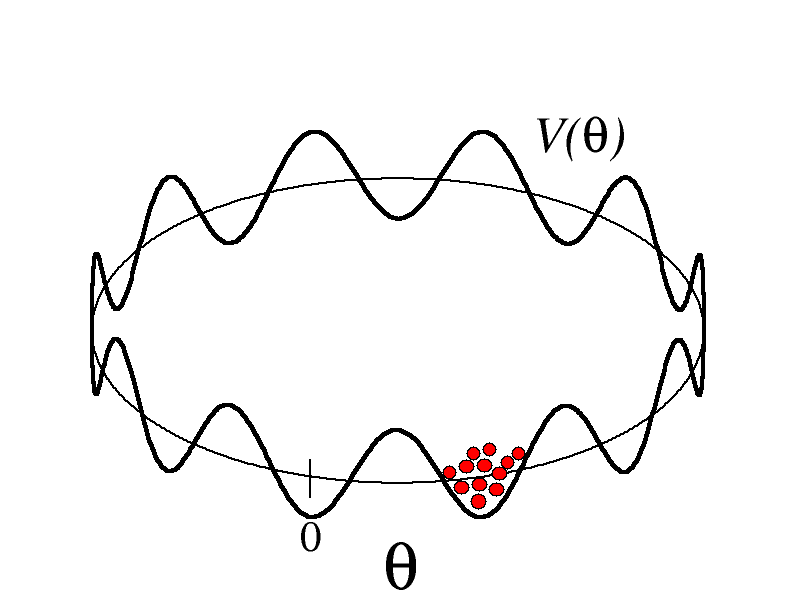}       
\includegraphics[width=0.4\columnwidth]{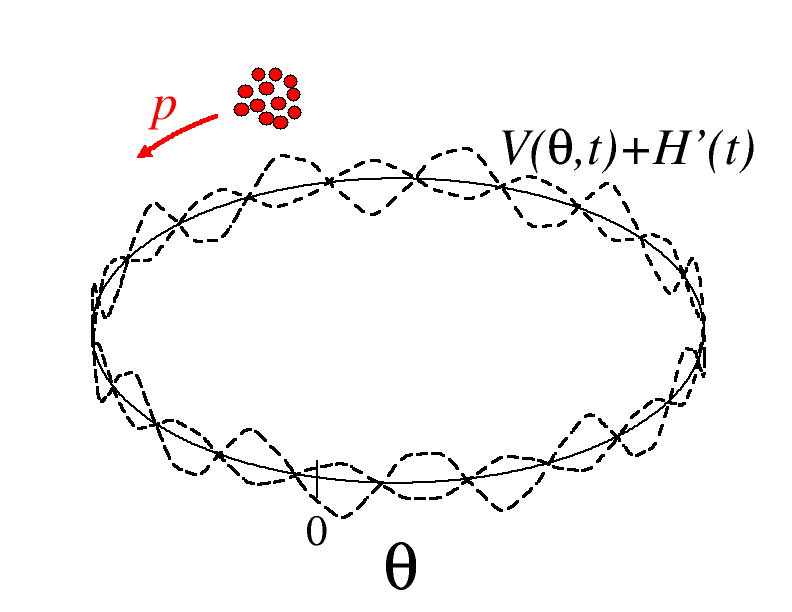}       
\caption{(color online) Left panel shows an initial stage of an experiment: ultra-cold atoms are prepared in a local minimum of the potential $V(\theta)=\lambda\cos(s\theta)$ in a toroidal trap. Then, the temporal modulation of the potential, $V(\theta,t)=\lambda\cos(s\theta)\cos(\omega t)$, and a temporal disorder $H'(t)$ are turned on and atoms are kicked, $p\approx m\omega/s$, so that the $s:1$ resonance condition is fulfilled --- see right panel. Monitoring atomic density at a fixed position versus time allows one to investigate localization properties of the system.}
\label{ring}   
\end{figure} 

Another interesting possibility is to use a toroidal trap where atoms are forced to move on a ring,  {while} the transverse degrees of freedom  {are} frozen by a tight confinement. Using suitable phase masks, it is possible to shape the transverse profile 
of a coherent laser beam and to produce a flexible angular dependence of the laser intensity~\cite{Wright2000,Wang2009}. When those beams are sent on atoms on a ring,  {they create} a tunable optical potential whose temporal dependence can also be controlled by the experimentalist. A simple $V(\theta)=\lambda\cos(s\theta)$ dependence is obtained using order $s$ Gauss-Laguerre modes \cite{Clifford1998}.  {The} ultra-cold atomic gas can be initially loaded in  {a single potential minimum}, see Fig.~\ref{ring} for schematic plot of an experiment.
The standard toolbox of ultra-cold atomic physics (kicks by laser fields, magnetic fields, microwave fields, etc.) can then be used to create an atomic cloud rotating in the toroidal trap and to modulate the optical potential at a convenient frequency, $V(\theta,t)=\lambda\cos(s\theta)\cos(\omega t)$,
in order to match the $s:1$ resonance. Then, the single-particle Hamiltonian, $H=\frac{p^2}{2m}+V(\theta,t)$, in the rotating frame $\Theta=\theta-\omega t/s$, can be approximated by the secular Hamiltonian $H\approx \frac{P^2}{2m}+\frac{\lambda}{2}\cos(s\Theta)$.  The advantage of this setup is that 
the interaction $U_{ii}$ is purely diagonal,
so that the effective Hamiltonian (\ref{tight_q}) is given by the standard Bose-Hubbard model. Adding, by means of conveniently driven intensity of the laser beam, a perturbation $H'(t)=g(\theta)f(t)$, where $g(\theta)$ is any regular function which consists of at least $s$ harmonics and $f(t)$ is a temporally disorder function, leads to the final system described by the Hamiltonian (\ref{htot}). Transport and localization properties are again easily probed by measuring the atomic density at a fixed position vs. time. 

To summarize we have considered periodically driven many-body systems. Time-independent systems with spatially periodic potentials are standard models of space crystals. It turns out that systems described by time periodic Hamiltonians can reveal crystalline properties in time if they are resonantly driven and if the $s:1$ resonance corresponds to $s\gg 1$. These systems constitute models of time crystals in the same sense as their spatially periodic counterparts are common models of space crystals \cite{Sacha2017review}. In the present 
 {work} we have focused on resonantly driven many-body systems in the presence of a temporal disorder and provided a possible scenarios to observe the many body localization in the time domain.


We are grateful to Dominique Delande and Kuba Zakrzewski for fruitful discussion. Support of the National Science Centre, Poland via projects 2016/23/B/ST3/00647 (MM), 2016/20/W/ST4/00314 (KG) and 2016/21/B/ST2/01095 (KS) is acknowledged. {This work was performed with the support of EU via Horizon2020 FET project QUIC (nr. 641122).}


\begin{thebibliography}{81}%
\makeatletter
\providecommand \@ifxundefined [1]{%
 \@ifx{#1\undefined}
}%
\providecommand \@ifnum [1]{%
 \ifnum #1\expandafter \@firstoftwo
 \else \expandafter \@secondoftwo
 \fi
}%
\providecommand \@ifx [1]{%
 \ifx #1\expandafter \@firstoftwo
 \else \expandafter \@secondoftwo
 \fi
}%
\providecommand \natexlab [1]{#1}%
\providecommand \enquote  [1]{``#1''}%
\providecommand \bibnamefont  [1]{#1}%
\providecommand \bibfnamefont [1]{#1}%
\providecommand \citenamefont [1]{#1}%
\providecommand \href@noop [0]{\@secondoftwo}%
\providecommand \href [0]{\begingroup \@sanitize@url \@href}%
\providecommand \@href[1]{\@@startlink{#1}\@@href}%
\providecommand \@@href[1]{\endgroup#1\@@endlink}%
\providecommand \@sanitize@url [0]{\catcode `\\12\catcode `\$12\catcode
  `\&12\catcode `\#12\catcode `\^12\catcode `\_12\catcode `\%12\relax}%
\providecommand \@@startlink[1]{}%
\providecommand \@@endlink[0]{}%
\providecommand \url  [0]{\begingroup\@sanitize@url \@url }%
\providecommand \@url [1]{\endgroup\@href {#1}{\urlprefix }}%
\providecommand \urlprefix  [0]{URL }%
\providecommand \Eprint [0]{\href }%
\providecommand \doibase [0]{http://dx.doi.org/}%
\providecommand \selectlanguage [0]{\@gobble}%
\providecommand \bibinfo  [0]{\@secondoftwo}%
\providecommand \bibfield  [0]{\@secondoftwo}%
\providecommand \translation [1]{[#1]}%
\providecommand \BibitemOpen [0]{}%
\providecommand \bibitemStop [0]{}%
\providecommand \bibitemNoStop [0]{.\EOS\space}%
\providecommand \EOS [0]{\spacefactor3000\relax}%
\providecommand \BibitemShut  [1]{\csname bibitem#1\endcsname}%
\let\auto@bib@innerbib\@empty
\bibitem [{\citenamefont {Basko}\ \emph {et~al.}(2006)\citenamefont {Basko},
  \citenamefont {Aleiner},\ and\ \citenamefont {Altshuler}}]{Basko06}%
  \BibitemOpen
  \bibfield  {author} {\bibinfo {author} {\bibfnamefont {D.~M.}\ \bibnamefont
  {Basko}}, \bibinfo {author} {\bibfnamefont {I.~L.}\ \bibnamefont {Aleiner}},
  \ and\ \bibinfo {author} {\bibfnamefont {B.~L.}\ \bibnamefont {Altshuler}},\
  }\href {\doibase 10.1016/j.aop.2005.11.014} {\bibfield  {journal} {\bibinfo
  {journal} {Ann. Phys.}\ }\textbf {\bibinfo {volume} {321}},\ \bibinfo {pages}
  {1126} (\bibinfo {year} {2006})}\BibitemShut {NoStop}%
\bibitem [{\citenamefont {Oganesyan}\ and\ \citenamefont
  {Huse}(2007)}]{Oganesyan07}%
  \BibitemOpen
  \bibfield  {author} {\bibinfo {author} {\bibfnamefont {V.}~\bibnamefont
  {Oganesyan}}\ and\ \bibinfo {author} {\bibfnamefont {D.~A.}\ \bibnamefont
  {Huse}},\ }\href {\doibase 10.1103/PhysRevB.75.155111} {\bibfield  {journal}
  {\bibinfo  {journal} {Phys. Rev. B}\ }\textbf {\bibinfo {volume} {75}},\
  \bibinfo {pages} {155111} (\bibinfo {year} {2007})}\BibitemShut {NoStop}%
\bibitem [{\citenamefont {\ifmmode \check{Z}\else
  \v{Z}\fi{}nidari\ifmmode~\check{c}\else \v{c}\fi{}}\ \emph
  {et~al.}(2008)\citenamefont {\ifmmode \check{Z}\else
  \v{Z}\fi{}nidari\ifmmode~\check{c}\else \v{c}\fi{}}, \citenamefont {Prosen},\
  and\ \citenamefont {Prelov\ifmmode~\check{s}\else
  \v{s}\fi{}ek}}]{Znidaric08}%
  \BibitemOpen
  \bibfield  {author} {\bibinfo {author} {\bibfnamefont {M.}~\bibnamefont
  {\ifmmode \check{Z}\else \v{Z}\fi{}nidari\ifmmode~\check{c}\else
  \v{c}\fi{}}}, \bibinfo {author} {\bibfnamefont {T.}~\bibnamefont {Prosen}}, \
  and\ \bibinfo {author} {\bibfnamefont {P.}~\bibnamefont
  {Prelov\ifmmode~\check{s}\else \v{s}\fi{}ek}},\ }\href {\doibase
  10.1103/PhysRevB.77.064426} {\bibfield  {journal} {\bibinfo  {journal} {Phys.
  Rev. B}\ }\textbf {\bibinfo {volume} {77}},\ \bibinfo {pages} {064426}
  (\bibinfo {year} {2008})}\BibitemShut {NoStop}%
\bibitem [{\citenamefont {Huse}\ \emph {et~al.}(2014)\citenamefont {Huse},
  \citenamefont {Nandkishore},\ and\ \citenamefont {Oganesyan}}]{Huse14}%
  \BibitemOpen
  \bibfield  {author} {\bibinfo {author} {\bibfnamefont {D.~A.}\ \bibnamefont
  {Huse}}, \bibinfo {author} {\bibfnamefont {R.}~\bibnamefont {Nandkishore}}, \
  and\ \bibinfo {author} {\bibfnamefont {V.}~\bibnamefont {Oganesyan}},\ }\href
  {\doibase 10.1103/PhysRevB.90.174202} {\bibfield  {journal} {\bibinfo
  {journal} {Phys. Rev. B}\ }\textbf {\bibinfo {volume} {90}},\ \bibinfo
  {pages} {174202} (\bibinfo {year} {2014})}\BibitemShut {NoStop}%
\bibitem [{\citenamefont {Nandkishore}\ and\ \citenamefont
  {Huse}(2015)}]{Rahul15}%
  \BibitemOpen
  \bibfield  {author} {\bibinfo {author} {\bibfnamefont {R.}~\bibnamefont
  {Nandkishore}}\ and\ \bibinfo {author} {\bibfnamefont {D.~A.}\ \bibnamefont
  {Huse}},\ }\href@noop {} {\bibfield  {journal} {\bibinfo  {journal} {Ann.
  Rev. Cond. Mat. Phys.}\ }\textbf {\bibinfo {volume} {6}},\ \bibinfo {pages}
  {15} (\bibinfo {year} {2015})}\BibitemShut {NoStop}%
\bibitem [{\citenamefont {Berkelbach}\ and\ \citenamefont
  {Reichman}(2010)}]{berkelbach10}%
  \BibitemOpen
  \bibfield  {author} {\bibinfo {author} {\bibfnamefont {T.~C.}\ \bibnamefont
  {Berkelbach}}\ and\ \bibinfo {author} {\bibfnamefont {D.~R.}\ \bibnamefont
  {Reichman}},\ }\href {\doibase 10.1103/PhysRevB.81.224429} {\bibfield
  {journal} {\bibinfo  {journal} {Phys. Rev. B}\ }\textbf {\bibinfo {volume}
  {81}},\ \bibinfo {pages} {224429} (\bibinfo {year} {2010})}\BibitemShut
  {NoStop}%
\bibitem [{\citenamefont {Bari\v{s}i\'{c}}\ and\ \citenamefont
  {Prelov\v{s}ek}(2010)}]{barisic10}%
  \BibitemOpen
  \bibfield  {author} {\bibinfo {author} {\bibfnamefont {O.~S.}\ \bibnamefont
  {Bari\v{s}i\'{c}}}\ and\ \bibinfo {author} {\bibfnamefont {P.}~\bibnamefont
  {Prelov\v{s}ek}},\ }\href {\doibase 10.1103/PhysRevB.82.161106} {\bibfield
  {journal} {\bibinfo  {journal} {Phys. Rev. B}\ }\textbf {\bibinfo {volume}
  {82}},\ \bibinfo {pages} {161106} (\bibinfo {year} {2010})}\BibitemShut
  {NoStop}%
\bibitem [{\citenamefont {Agarwal}\ \emph {et~al.}(2015)\citenamefont
  {Agarwal}, \citenamefont {Gopalakrishnan}, \citenamefont {Knap},
  \citenamefont {M\"{u}ller},\ and\ \citenamefont {Demler}}]{agarwal15}%
  \BibitemOpen
  \bibfield  {author} {\bibinfo {author} {\bibfnamefont {K.}~\bibnamefont
  {Agarwal}}, \bibinfo {author} {\bibfnamefont {S.}~\bibnamefont
  {Gopalakrishnan}}, \bibinfo {author} {\bibfnamefont {M.}~\bibnamefont
  {Knap}}, \bibinfo {author} {\bibfnamefont {M.}~\bibnamefont {M\"{u}ller}}, \
  and\ \bibinfo {author} {\bibfnamefont {E.}~\bibnamefont {Demler}},\ }\href
  {\doibase 10.1103/PhysRevLett.114.160401} {\bibfield  {journal} {\bibinfo
  {journal} {Phys. Rev. Lett.}\ }\textbf {\bibinfo {volume} {114}},\ \bibinfo
  {pages} {160401} (\bibinfo {year} {2015})}\BibitemShut {NoStop}%
\bibitem [{\citenamefont {Bar~Lev}\ \emph {et~al.}(2015)\citenamefont
  {Bar~Lev}, \citenamefont {Cohen},\ and\ \citenamefont {Reichman}}]{lev15}%
  \BibitemOpen
  \bibfield  {author} {\bibinfo {author} {\bibfnamefont {Y.}~\bibnamefont
  {Bar~Lev}}, \bibinfo {author} {\bibfnamefont {G.}~\bibnamefont {Cohen}}, \
  and\ \bibinfo {author} {\bibfnamefont {D.~R.}\ \bibnamefont {Reichman}},\
  }\href {\doibase 10.1103/PhysRevLett.114.100601} {\bibfield  {journal}
  {\bibinfo  {journal} {Phys. Rev. Lett.}\ }\textbf {\bibinfo {volume} {114}},\
  \bibinfo {pages} {100601} (\bibinfo {year} {2015})}\BibitemShut {NoStop}%
\bibitem [{\citenamefont {Steinigeweg}\ \emph {et~al.}(2016)\citenamefont
  {Steinigeweg}, \citenamefont {Herbrych}, \citenamefont {Pollmann},\ and\
  \citenamefont {Brenig}}]{steinigeweg15}%
  \BibitemOpen
  \bibfield  {author} {\bibinfo {author} {\bibfnamefont {R.}~\bibnamefont
  {Steinigeweg}}, \bibinfo {author} {\bibfnamefont {J.}~\bibnamefont
  {Herbrych}}, \bibinfo {author} {\bibfnamefont {F.}~\bibnamefont {Pollmann}},
  \ and\ \bibinfo {author} {\bibfnamefont {W.}~\bibnamefont {Brenig}},\ }\href
  {\doibase 10.1103/PhysRevB.94.180401} {\bibfield  {journal} {\bibinfo
  {journal} {Phys. Rev. B}\ }\textbf {\bibinfo {volume} {94}},\ \bibinfo
  {pages} {180401(R)} (\bibinfo {year} {2016})}\BibitemShut {NoStop}%
\bibitem [{\citenamefont {Bari\v{s}i\'{c}}\ \emph {et~al.}(2016)\citenamefont
  {Bari\v{s}i\'{c}}, \citenamefont {Kokalj}, \citenamefont {Balog},\ and\
  \citenamefont {Prelov\v{s}ek}}]{barisic16}%
  \BibitemOpen
  \bibfield  {author} {\bibinfo {author} {\bibfnamefont {O.~S.}\ \bibnamefont
  {Bari\v{s}i\'{c}}}, \bibinfo {author} {\bibfnamefont {J.}~\bibnamefont
  {Kokalj}}, \bibinfo {author} {\bibfnamefont {I.}~\bibnamefont {Balog}}, \
  and\ \bibinfo {author} {\bibfnamefont {P.}~\bibnamefont {Prelov\v{s}ek}},\
  }\href {\doibase 10.1103/PhysRevB.94.045126} {\bibfield  {journal} {\bibinfo
  {journal} {Phys. Rev. B}\ }\textbf {\bibinfo {volume} {94}},\ \bibinfo
  {pages} {045126} (\bibinfo {year} {2016})}\BibitemShut {NoStop}%
\bibitem [{\citenamefont {Kozarzewski}\ \emph {et~al.}(2016)\citenamefont
  {Kozarzewski}, \citenamefont {Prelov\v{s}ek},\ and\ \citenamefont
  {Mierzejewski}}]{kozarzewski16}%
  \BibitemOpen
  \bibfield  {author} {\bibinfo {author} {\bibfnamefont {M.}~\bibnamefont
  {Kozarzewski}}, \bibinfo {author} {\bibfnamefont {P.}~\bibnamefont
  {Prelov\v{s}ek}}, \ and\ \bibinfo {author} {\bibfnamefont {M.}~\bibnamefont
  {Mierzejewski}},\ }\href {\doibase 10.1103/PhysRevB.93.235151} {\bibfield
  {journal} {\bibinfo  {journal} {Phys. Rev. B}\ }\textbf {\bibinfo {volume}
  {93}},\ \bibinfo {pages} {235151} (\bibinfo {year} {2016})}\BibitemShut
  {NoStop}%
\bibitem [{\citenamefont {Pal}\ and\ \citenamefont {Huse}(2010)}]{pal10}%
  \BibitemOpen
  \bibfield  {author} {\bibinfo {author} {\bibfnamefont {A.}~\bibnamefont
  {Pal}}\ and\ \bibinfo {author} {\bibfnamefont {D.~A.}\ \bibnamefont {Huse}},\
  }\href {\doibase 10.1103/PhysRevB.82.174411} {\bibfield  {journal} {\bibinfo
  {journal} {Phys. Rev. B}\ }\textbf {\bibinfo {volume} {82}},\ \bibinfo
  {pages} {174411} (\bibinfo {year} {2010})}\BibitemShut {NoStop}%
\bibitem [{\citenamefont {Serbyn}\ \emph
  {et~al.}(2013{\natexlab{a}})\citenamefont {Serbyn}, \citenamefont
  {Papi\ifmmode~\acute{c}\else \'{c}\fi{}},\ and\ \citenamefont
  {Abanin}}]{serbyn2013}%
  \BibitemOpen
  \bibfield  {author} {\bibinfo {author} {\bibfnamefont {M.}~\bibnamefont
  {Serbyn}}, \bibinfo {author} {\bibfnamefont {Z.}~\bibnamefont
  {Papi\ifmmode~\acute{c}\else \'{c}\fi{}}}, \ and\ \bibinfo {author}
  {\bibfnamefont {D.~A.}\ \bibnamefont {Abanin}},\ }\href {\doibase
  10.1103/PhysRevLett.111.127201} {\bibfield  {journal} {\bibinfo  {journal}
  {Phys. Rev. Lett.}\ }\textbf {\bibinfo {volume} {111}},\ \bibinfo {pages}
  {127201} (\bibinfo {year} {2013}{\natexlab{a}})}\BibitemShut {NoStop}%
\bibitem [{\citenamefont {Bar~Lev}\ and\ \citenamefont
  {Reichman}(2014)}]{lev14}%
  \BibitemOpen
  \bibfield  {author} {\bibinfo {author} {\bibfnamefont {Y.}~\bibnamefont
  {Bar~Lev}}\ and\ \bibinfo {author} {\bibfnamefont {D.~R.}\ \bibnamefont
  {Reichman}},\ }\href {\doibase 10.1103/PhysRevB.89.220201} {\bibfield
  {journal} {\bibinfo  {journal} {Phys. Rev. B}\ }\textbf {\bibinfo {volume}
  {89}},\ \bibinfo {pages} {220201} (\bibinfo {year} {2014})}\BibitemShut
  {NoStop}%
\bibitem [{\citenamefont {Schreiber}\ \emph {et~al.}(2015)\citenamefont
  {Schreiber}, \citenamefont {Hodgman}, \citenamefont {Bordia}, \citenamefont
  {L\"{u}schen}, \citenamefont {Fischer}, \citenamefont {Vosk}, \citenamefont
  {Altman}, \citenamefont {Schneider},\ and\ \citenamefont
  {Bloch}}]{schreiber15}%
  \BibitemOpen
  \bibfield  {author} {\bibinfo {author} {\bibfnamefont {M.}~\bibnamefont
  {Schreiber}}, \bibinfo {author} {\bibfnamefont {S.~S.}\ \bibnamefont
  {Hodgman}}, \bibinfo {author} {\bibfnamefont {P.}~\bibnamefont {Bordia}},
  \bibinfo {author} {\bibfnamefont {H.~P.}\ \bibnamefont {L\"{u}schen}},
  \bibinfo {author} {\bibfnamefont {M.~H.}\ \bibnamefont {Fischer}}, \bibinfo
  {author} {\bibfnamefont {R.}~\bibnamefont {Vosk}}, \bibinfo {author}
  {\bibfnamefont {E.}~\bibnamefont {Altman}}, \bibinfo {author} {\bibfnamefont
  {U.}~\bibnamefont {Schneider}}, \ and\ \bibinfo {author} {\bibfnamefont
  {I.}~\bibnamefont {Bloch}},\ }\href {\doibase 10.1126/science.aaa7432}
  {\bibfield  {journal} {\bibinfo  {journal} {Science}\ }\textbf {\bibinfo
  {volume} {349}},\ \bibinfo {pages} {842} (\bibinfo {year}
  {2015})}\BibitemShut {NoStop}%
\bibitem [{\citenamefont {Serbyn}\ \emph {et~al.}(2015)\citenamefont {Serbyn},
  \citenamefont {Papi{\'{c}}},\ and\ \citenamefont {Abanin}}]{serbyn15}%
  \BibitemOpen
  \bibfield  {author} {\bibinfo {author} {\bibfnamefont {M.}~\bibnamefont
  {Serbyn}}, \bibinfo {author} {\bibfnamefont {Z.}~\bibnamefont {Papi{\'{c}}}},
  \ and\ \bibinfo {author} {\bibfnamefont {D.~A.}\ \bibnamefont {Abanin}},\
  }\href {\doibase 10.1103/PhysRevX.5.041047} {\bibfield  {journal} {\bibinfo
  {journal} {Phys. Rev. X}\ }\textbf {\bibinfo {volume} {5}},\ \bibinfo {pages}
  {041047} (\bibinfo {year} {2015})}\BibitemShut {NoStop}%
\bibitem [{\citenamefont {Khemani}\ \emph {et~al.}(2015)\citenamefont
  {Khemani}, \citenamefont {Nandkishore},\ and\ \citenamefont
  {Sondhi}}]{khemani15}%
  \BibitemOpen
  \bibfield  {author} {\bibinfo {author} {\bibfnamefont {V.}~\bibnamefont
  {Khemani}}, \bibinfo {author} {\bibfnamefont {R.}~\bibnamefont
  {Nandkishore}}, \ and\ \bibinfo {author} {\bibfnamefont {S.~L.}\ \bibnamefont
  {Sondhi}},\ }\href {\doibase 10.1038/nphys3344} {\bibfield  {journal}
  {\bibinfo  {journal} {Nat. Phys.}\ }\textbf {\bibinfo {volume} {11}},\
  \bibinfo {pages} {560} (\bibinfo {year} {2015})}\BibitemShut {NoStop}%
\bibitem [{\citenamefont {Luitz}\ \emph
  {et~al.}(2015{\natexlab{a}})\citenamefont {Luitz}, \citenamefont
  {Laflorencie},\ and\ \citenamefont {Alet}}]{luitz16}%
  \BibitemOpen
  \bibfield  {author} {\bibinfo {author} {\bibfnamefont {D.~J.}\ \bibnamefont
  {Luitz}}, \bibinfo {author} {\bibfnamefont {N.}~\bibnamefont {Laflorencie}},
  \ and\ \bibinfo {author} {\bibfnamefont {F.}~\bibnamefont {Alet}},\ }\href
  {\doibase 10.1103/PhysRevB.93.060201} {\bibfield  {journal} {\bibinfo
  {journal} {Phys. Rev. B}\ }\textbf {\bibinfo {volume} {93}},\ \bibinfo
  {pages} {060201} (\bibinfo {year} {2015}{\natexlab{a}})}\BibitemShut
  {NoStop}%
\bibitem [{\citenamefont {Gornyi}\ \emph {et~al.}(2005)\citenamefont {Gornyi},
  \citenamefont {Mirlin},\ and\ \citenamefont {Polyakov}}]{gornyi05}%
  \BibitemOpen
  \bibfield  {author} {\bibinfo {author} {\bibfnamefont {I.~V.}\ \bibnamefont
  {Gornyi}}, \bibinfo {author} {\bibfnamefont {A.~D.}\ \bibnamefont {Mirlin}},
  \ and\ \bibinfo {author} {\bibfnamefont {D.~G.}\ \bibnamefont {Polyakov}},\
  }\href {\doibase 10.1103/PhysRevLett.95.206603} {\bibfield  {journal}
  {\bibinfo  {journal} {Phys. Rev. Lett.}\ }\textbf {\bibinfo {volume} {95}},\
  \bibinfo {pages} {206603} (\bibinfo {year} {2005})}\BibitemShut {NoStop}%
\bibitem [{\citenamefont {Altman}\ and\ \citenamefont {Vosk}(2015)}]{altman15}%
  \BibitemOpen
  \bibfield  {author} {\bibinfo {author} {\bibfnamefont {E.}~\bibnamefont
  {Altman}}\ and\ \bibinfo {author} {\bibfnamefont {R.}~\bibnamefont {Vosk}},\
  }\href {\doibase 10.1146/annurev-conmatphys-031214-014701} {\bibfield
  {journal} {\bibinfo  {journal} {Annu. Rev. Condens. Matter Phys.}\ }\textbf
  {\bibinfo {volume} {6}},\ \bibinfo {pages} {383} (\bibinfo {year}
  {2015})}\BibitemShut {NoStop}%
\bibitem [{\citenamefont {De~Luca}\ and\ \citenamefont
  {Scardicchio}(2013)}]{deluca13}%
  \BibitemOpen
  \bibfield  {author} {\bibinfo {author} {\bibfnamefont {A.}~\bibnamefont
  {De~Luca}}\ and\ \bibinfo {author} {\bibfnamefont {A.}~\bibnamefont
  {Scardicchio}},\ }\href {http://stacks.iop.org/0295-5075/101/i=3/a=37003}
  {\bibfield  {journal} {\bibinfo  {journal} {EPL (Europhysics Letters)}\
  }\textbf {\bibinfo {volume} {101}},\ \bibinfo {pages} {37003} (\bibinfo
  {year} {2013})}\BibitemShut {NoStop}%
\bibitem [{\citenamefont {Gramsch}\ and\ \citenamefont
  {Rigol}(2012)}]{gramsch12}%
  \BibitemOpen
  \bibfield  {author} {\bibinfo {author} {\bibfnamefont {C.}~\bibnamefont
  {Gramsch}}\ and\ \bibinfo {author} {\bibfnamefont {M.}~\bibnamefont
  {Rigol}},\ }\href {\doibase 10.1103/PhysRevA.86.053615} {\bibfield  {journal}
  {\bibinfo  {journal} {Phys. Rev. A}\ }\textbf {\bibinfo {volume} {86}},\
  \bibinfo {pages} {053615} (\bibinfo {year} {2012})}\BibitemShut {NoStop}%
\bibitem [{\citenamefont {De~Luca}\ \emph {et~al.}(2014)\citenamefont
  {De~Luca}, \citenamefont {Altshuler}, \citenamefont {Kravtsov},\ and\
  \citenamefont {Scardicchio}}]{deluca14}%
  \BibitemOpen
  \bibfield  {author} {\bibinfo {author} {\bibfnamefont {A.}~\bibnamefont
  {De~Luca}}, \bibinfo {author} {\bibfnamefont {B.~L.}\ \bibnamefont
  {Altshuler}}, \bibinfo {author} {\bibfnamefont {V.~E.}\ \bibnamefont
  {Kravtsov}}, \ and\ \bibinfo {author} {\bibfnamefont {A.}~\bibnamefont
  {Scardicchio}},\ }\href {\doibase 10.1103/PhysRevLett.113.046806} {\bibfield
  {journal} {\bibinfo  {journal} {Phys. Rev. Lett.}\ }\textbf {\bibinfo
  {volume} {113}},\ \bibinfo {pages} {046806} (\bibinfo {year}
  {2014})}\BibitemShut {NoStop}%
\bibitem [{\citenamefont {Huse}\ \emph {et~al.}(2013)\citenamefont {Huse},
  \citenamefont {Nandkishore}, \citenamefont {Oganesyan}, \citenamefont {Pal},\
  and\ \citenamefont {Sondhi}}]{huse13}%
  \BibitemOpen
  \bibfield  {author} {\bibinfo {author} {\bibfnamefont {D.~A.}\ \bibnamefont
  {Huse}}, \bibinfo {author} {\bibfnamefont {R.}~\bibnamefont {Nandkishore}},
  \bibinfo {author} {\bibfnamefont {V.}~\bibnamefont {Oganesyan}}, \bibinfo
  {author} {\bibfnamefont {A.}~\bibnamefont {Pal}}, \ and\ \bibinfo {author}
  {\bibfnamefont {S.~L.}\ \bibnamefont {Sondhi}},\ }\href {\doibase
  10.1103/PhysRevB.88.014206} {\bibfield  {journal} {\bibinfo  {journal} {Phys.
  Rev. B}\ }\textbf {\bibinfo {volume} {88}},\ \bibinfo {pages} {014206}
  (\bibinfo {year} {2013})}\BibitemShut {NoStop}%
\bibitem [{\citenamefont {Rademaker}\ and\ \citenamefont
  {Ortu\~no}(2016)}]{rademaker16}%
  \BibitemOpen
  \bibfield  {author} {\bibinfo {author} {\bibfnamefont {L.}~\bibnamefont
  {Rademaker}}\ and\ \bibinfo {author} {\bibfnamefont {M.}~\bibnamefont
  {Ortu\~no}},\ }\href {\doibase 10.1103/PhysRevLett.116.010404} {\bibfield
  {journal} {\bibinfo  {journal} {Phys. Rev. Lett.}\ }\textbf {\bibinfo
  {volume} {116}},\ \bibinfo {pages} {010404} (\bibinfo {year}
  {2016})}\BibitemShut {NoStop}%
\bibitem [{\citenamefont {Chandran}\ \emph {et~al.}(2015)\citenamefont
  {Chandran}, \citenamefont {Kim}, \citenamefont {Vidal},\ and\ \citenamefont
  {Abanin}}]{chandran15}%
  \BibitemOpen
  \bibfield  {author} {\bibinfo {author} {\bibfnamefont {A.}~\bibnamefont
  {Chandran}}, \bibinfo {author} {\bibfnamefont {I.~H.}\ \bibnamefont {Kim}},
  \bibinfo {author} {\bibfnamefont {G.}~\bibnamefont {Vidal}}, \ and\ \bibinfo
  {author} {\bibfnamefont {D.~A.}\ \bibnamefont {Abanin}},\ }\href {\doibase
  10.1103/PhysRevB.91.085425} {\bibfield  {journal} {\bibinfo  {journal} {Phys.
  Rev. B}\ }\textbf {\bibinfo {volume} {91}},\ \bibinfo {pages} {085425}
  (\bibinfo {year} {2015})}\BibitemShut {NoStop}%
\bibitem [{\citenamefont {Ros}\ \emph {et~al.}(2015)\citenamefont {Ros},
  \citenamefont {M{\"u}ller},\ and\ \citenamefont {Scardicchio}}]{ros15}%
  \BibitemOpen
  \bibfield  {author} {\bibinfo {author} {\bibfnamefont {V.}~\bibnamefont
  {Ros}}, \bibinfo {author} {\bibfnamefont {M.}~\bibnamefont {M{\"u}ller}}, \
  and\ \bibinfo {author} {\bibfnamefont {A.}~\bibnamefont {Scardicchio}},\
  }\href {\doibase http://dx.doi.org/10.1016/j.nuclphysb.2014.12.014}
  {\bibfield  {journal} {\bibinfo  {journal} {Nuclear Physics B}\ }\textbf
  {\bibinfo {volume} {891}},\ \bibinfo {pages} {420} (\bibinfo {year}
  {2015})}\BibitemShut {NoStop}%
\bibitem [{\citenamefont {Eisert}\ \emph {et~al.}(2015)\citenamefont {Eisert},
  \citenamefont {Friesdorf},\ and\ \citenamefont {Gogolin}}]{eisert15}%
  \BibitemOpen
  \bibfield  {author} {\bibinfo {author} {\bibfnamefont {J.}~\bibnamefont
  {Eisert}}, \bibinfo {author} {\bibfnamefont {M.}~\bibnamefont {Friesdorf}}, \
  and\ \bibinfo {author} {\bibfnamefont {C.}~\bibnamefont {Gogolin}},\ }\href
  {http://dx.doi.org/10.1038/nphys3215} {\bibfield  {journal} {\bibinfo
  {journal} {Nat Phys}\ }\textbf {\bibinfo {volume} {11}},\ \bibinfo {pages}
  {124} (\bibinfo {year} {2015})}\BibitemShut {NoStop}%
\bibitem [{\citenamefont {Serbyn}\ \emph {et~al.}(2014)\citenamefont {Serbyn},
  \citenamefont {Knap}, \citenamefont {Gopalakrishnan}, \citenamefont
  {Papi\'{c}}, \citenamefont {Yao}, \citenamefont {Laumann}, \citenamefont
  {Abanin}, \citenamefont {Lukin},\ and\ \citenamefont {Demler}}]{serbyn141}%
  \BibitemOpen
  \bibfield  {author} {\bibinfo {author} {\bibfnamefont {M.}~\bibnamefont
  {Serbyn}}, \bibinfo {author} {\bibfnamefont {M.}~\bibnamefont {Knap}},
  \bibinfo {author} {\bibfnamefont {S.}~\bibnamefont {Gopalakrishnan}},
  \bibinfo {author} {\bibfnamefont {Z.}~\bibnamefont {Papi\'{c}}}, \bibinfo
  {author} {\bibfnamefont {N.~Y.}\ \bibnamefont {Yao}}, \bibinfo {author}
  {\bibfnamefont {C.~R.}\ \bibnamefont {Laumann}}, \bibinfo {author}
  {\bibfnamefont {D.~A.}\ \bibnamefont {Abanin}}, \bibinfo {author}
  {\bibfnamefont {M.~D.}\ \bibnamefont {Lukin}}, \ and\ \bibinfo {author}
  {\bibfnamefont {E.~A.}\ \bibnamefont {Demler}},\ }\href {\doibase
  10.1103/PhysRevLett.113.147204} {\bibfield  {journal} {\bibinfo  {journal}
  {Phys. Rev. Lett.}\ }\textbf {\bibinfo {volume} {113}},\ \bibinfo {pages}
  {147204} (\bibinfo {year} {2014})}\BibitemShut {NoStop}%
\bibitem [{\citenamefont {Sierant}\ \emph {et~al.}(2017)\citenamefont
  {Sierant}, \citenamefont {Delande},\ and\ \citenamefont
  {Zakrzewski}}]{zakrzewski16}%
  \BibitemOpen
  \bibfield  {author} {\bibinfo {author} {\bibfnamefont {P.}~\bibnamefont
  {Sierant}}, \bibinfo {author} {\bibfnamefont {D.}~\bibnamefont {Delande}}, \
  and\ \bibinfo {author} {\bibfnamefont {J.}~\bibnamefont {Zakrzewski}},\
  }\href {\doibase 10.1103/PhysRevA.95.021601} {\bibfield  {journal} {\bibinfo
  {journal} {Phys. Rev. A}\ }\textbf {\bibinfo {volume} {95}},\ \bibinfo
  {pages} {021601} (\bibinfo {year} {2017})}\BibitemShut {NoStop}%
\bibitem [{\citenamefont {Monthus}\ and\ \citenamefont
  {Garel}(2010)}]{monthus10}%
  \BibitemOpen
  \bibfield  {author} {\bibinfo {author} {\bibfnamefont {C.}~\bibnamefont
  {Monthus}}\ and\ \bibinfo {author} {\bibfnamefont {T.}~\bibnamefont
  {Garel}},\ }\href {\doibase 10.1103/PhysRevB.81.134202} {\bibfield  {journal}
  {\bibinfo  {journal} {Phys. Rev. B}\ }\textbf {\bibinfo {volume} {81}},\
  \bibinfo {pages} {134202} (\bibinfo {year} {2010})}\BibitemShut {NoStop}%
\bibitem [{\citenamefont {Mierzejewski}\ \emph {et~al.}(2016)\citenamefont
  {Mierzejewski}, \citenamefont {Herbrych},\ and\ \citenamefont
  {Prelov{\v{s}}ek}}]{mierzejewski16}%
  \BibitemOpen
  \bibfield  {author} {\bibinfo {author} {\bibfnamefont {M.}~\bibnamefont
  {Mierzejewski}}, \bibinfo {author} {\bibfnamefont {J.}~\bibnamefont
  {Herbrych}}, \ and\ \bibinfo {author} {\bibfnamefont {P.}~\bibnamefont
  {Prelov{\v{s}}ek}},\ }\href {\doibase 10.1103/PhysRevB.94.224207} {\bibfield
  {journal} {\bibinfo  {journal} {Physical Review B}\ }\textbf {\bibinfo
  {volume} {94}},\ \bibinfo {pages} {224207} (\bibinfo {year}
  {2016})}\BibitemShut {NoStop}%
\bibitem [{\citenamefont {Prelov{\v{s}}ek}\ and\ \citenamefont
  {Herbrych}(2016)}]{prelovsek16}%
  \BibitemOpen
  \bibfield  {author} {\bibinfo {author} {\bibfnamefont {P.}~\bibnamefont
  {Prelov{\v{s}}ek}}\ and\ \bibinfo {author} {\bibfnamefont {J.}~\bibnamefont
  {Herbrych}},\ }\href {http://arxiv.org/abs/1609.05450} {\bibfield  {journal}
  {\bibinfo  {journal} {ArXiv e-prints}\ } (\bibinfo {year} {2016})},\ \Eprint
  {http://arxiv.org/abs/1609.05450} {1609.05450 [cond-mat.stat-mech]}
  \BibitemShut {NoStop}%
\bibitem [{\citenamefont {Mondaini}\ and\ \citenamefont
  {Rigol}(2015)}]{mondaini15}%
  \BibitemOpen
  \bibfield  {author} {\bibinfo {author} {\bibfnamefont {R.}~\bibnamefont
  {Mondaini}}\ and\ \bibinfo {author} {\bibfnamefont {M.}~\bibnamefont
  {Rigol}},\ }\href {\doibase 10.1103/PhysRevA.92.041601} {\bibfield  {journal}
  {\bibinfo  {journal} {Phys. Rev. A}\ }\textbf {\bibinfo {volume} {92}},\
  \bibinfo {pages} {041601(R)} (\bibinfo {year} {2015})}\BibitemShut {NoStop}%
\bibitem [{\citenamefont {Bardarson}\ \emph {et~al.}(2012)\citenamefont
  {Bardarson}, \citenamefont {Pollmann},\ and\ \citenamefont
  {Moore}}]{bardarson12}%
  \BibitemOpen
  \bibfield  {author} {\bibinfo {author} {\bibfnamefont {J.~H.}\ \bibnamefont
  {Bardarson}}, \bibinfo {author} {\bibfnamefont {F.}~\bibnamefont {Pollmann}},
  \ and\ \bibinfo {author} {\bibfnamefont {J.~E.}\ \bibnamefont {Moore}},\
  }\href {\doibase 10.1103/PhysRevLett.109.017202} {\bibfield  {journal}
  {\bibinfo  {journal} {Phys. Rev. Lett.}\ }\textbf {\bibinfo {volume} {109}},\
  \bibinfo {pages} {017202} (\bibinfo {year} {2012})}\BibitemShut {NoStop}%
\bibitem [{\citenamefont {Kj\"{a}ll}\ \emph {et~al.}(2014)\citenamefont
  {Kj\"{a}ll}, \citenamefont {Bardarson},\ and\ \citenamefont
  {Pollmann}}]{kjall14}%
  \BibitemOpen
  \bibfield  {author} {\bibinfo {author} {\bibfnamefont {J.~A.}\ \bibnamefont
  {Kj\"{a}ll}}, \bibinfo {author} {\bibfnamefont {J.~H.}\ \bibnamefont
  {Bardarson}}, \ and\ \bibinfo {author} {\bibfnamefont {F.}~\bibnamefont
  {Pollmann}},\ }\href {\doibase 10.1103/PhysRevLett.113.107204} {\bibfield
  {journal} {\bibinfo  {journal} {Phys. Rev. Lett.}\ }\textbf {\bibinfo
  {volume} {113}},\ \bibinfo {pages} {107204} (\bibinfo {year}
  {2014})}\BibitemShut {NoStop}%
\bibitem [{\citenamefont {Choi}\ \emph {et~al.}(2016)\citenamefont {Choi},
  \citenamefont {Hild}, \citenamefont {Zeiher}, \citenamefont {Schau{\ss}},
  \citenamefont {Rubio-Abadal}, \citenamefont {Yefsah}, \citenamefont
  {Khemani}, \citenamefont {Huse}, \citenamefont {Bloch},\ and\ \citenamefont
  {Gross}}]{choi16}%
  \BibitemOpen
  \bibfield  {author} {\bibinfo {author} {\bibfnamefont {J.-Y.}\ \bibnamefont
  {Choi}}, \bibinfo {author} {\bibfnamefont {S.}~\bibnamefont {Hild}}, \bibinfo
  {author} {\bibfnamefont {J.}~\bibnamefont {Zeiher}}, \bibinfo {author}
  {\bibfnamefont {P.}~\bibnamefont {Schau{\ss}}}, \bibinfo {author}
  {\bibfnamefont {A.}~\bibnamefont {Rubio-Abadal}}, \bibinfo {author}
  {\bibfnamefont {T.}~\bibnamefont {Yefsah}}, \bibinfo {author} {\bibfnamefont
  {V.}~\bibnamefont {Khemani}}, \bibinfo {author} {\bibfnamefont {D.~A.}\
  \bibnamefont {Huse}}, \bibinfo {author} {\bibfnamefont {I.}~\bibnamefont
  {Bloch}}, \ and\ \bibinfo {author} {\bibfnamefont {C.}~\bibnamefont
  {Gross}},\ }\href {\doibase 10.1126/science.aaf8834} {\bibfield  {journal}
  {\bibinfo  {journal} {Science}\ }\textbf {\bibinfo {volume} {352}},\ \bibinfo
  {pages} {1547} (\bibinfo {year} {2016})}\BibitemShut {NoStop}%
\bibitem [{\citenamefont {Bordia}\ \emph {et~al.}(2016)\citenamefont {Bordia},
  \citenamefont {L\"{u}schen}, \citenamefont {Hodgman}, \citenamefont
  {Schreiber}, \citenamefont {Bloch},\ and\ \citenamefont
  {Schneider}}]{bordia16}%
  \BibitemOpen
  \bibfield  {author} {\bibinfo {author} {\bibfnamefont {P.}~\bibnamefont
  {Bordia}}, \bibinfo {author} {\bibfnamefont {H.~P.}\ \bibnamefont
  {L\"{u}schen}}, \bibinfo {author} {\bibfnamefont {S.~S.}\ \bibnamefont
  {Hodgman}}, \bibinfo {author} {\bibfnamefont {M.}~\bibnamefont {Schreiber}},
  \bibinfo {author} {\bibfnamefont {I.}~\bibnamefont {Bloch}}, \ and\ \bibinfo
  {author} {\bibfnamefont {U.}~\bibnamefont {Schneider}},\ }\href {\doibase
  10.1103/PhysRevLett.116.140401} {\bibfield  {journal} {\bibinfo  {journal}
  {Phys. Rev. Lett.}\ }\textbf {\bibinfo {volume} {116}},\ \bibinfo {pages}
  {140401} (\bibinfo {year} {2016})}\BibitemShut {NoStop}%
\bibitem [{\citenamefont {{Bordia}}\ \emph {et~al.}(2017)\citenamefont
  {{Bordia}}, \citenamefont {{L{\"u}schen}}, \citenamefont {{Scherg}},
  \citenamefont {{Gopalakrishnan}}, \citenamefont {{Knap}}, \citenamefont
  {{Schneider}},\ and\ \citenamefont {{Bloch}}}]{bordia2017_1}%
  \BibitemOpen
  \bibfield  {author} {\bibinfo {author} {\bibfnamefont {P.}~\bibnamefont
  {{Bordia}}}, \bibinfo {author} {\bibfnamefont {H.}~\bibnamefont
  {{L{\"u}schen}}}, \bibinfo {author} {\bibfnamefont {S.}~\bibnamefont
  {{Scherg}}}, \bibinfo {author} {\bibfnamefont {S.}~\bibnamefont
  {{Gopalakrishnan}}}, \bibinfo {author} {\bibfnamefont {M.}~\bibnamefont
  {{Knap}}}, \bibinfo {author} {\bibfnamefont {U.}~\bibnamefont {{Schneider}}},
  \ and\ \bibinfo {author} {\bibfnamefont {I.}~\bibnamefont {{Bloch}}},\
  }\href@noop {} {\bibfield  {journal} {\bibinfo  {journal} {ArXiv e-prints}\ }
  (\bibinfo {year} {2017})},\ \Eprint {http://arxiv.org/abs/1704.03063}
  {arXiv:1704.03063 [cond-mat.quant-gas]} \BibitemShut {NoStop}%
\bibitem [{\citenamefont {Smith}\ \emph {et~al.}(2016)\citenamefont {Smith},
  \citenamefont {Lee}, \citenamefont {Richerme}, \citenamefont {Neyenhuis},
  \citenamefont {Hess}, \citenamefont {Hauke}, \citenamefont {Heyl},
  \citenamefont {Huse},\ and\ \citenamefont {Monroe}}]{smith2016}%
  \BibitemOpen
  \bibfield  {author} {\bibinfo {author} {\bibfnamefont {J.}~\bibnamefont
  {Smith}}, \bibinfo {author} {\bibfnamefont {A.}~\bibnamefont {Lee}}, \bibinfo
  {author} {\bibfnamefont {P.}~\bibnamefont {Richerme}}, \bibinfo {author}
  {\bibfnamefont {B.}~\bibnamefont {Neyenhuis}}, \bibinfo {author}
  {\bibfnamefont {P.~W.}\ \bibnamefont {Hess}}, \bibinfo {author}
  {\bibfnamefont {P.}~\bibnamefont {Hauke}}, \bibinfo {author} {\bibfnamefont
  {M.}~\bibnamefont {Heyl}}, \bibinfo {author} {\bibfnamefont {D.~A.}\
  \bibnamefont {Huse}}, \ and\ \bibinfo {author} {\bibfnamefont
  {C.}~\bibnamefont {Monroe}},\ }\href {\doibase dx.doi.org/10.1038/nphys3783}
  {\bibfield  {journal} {\bibinfo  {journal} {Nat. Phys.}\ }\textbf {\bibinfo
  {volume} {12}},\ \bibinfo {pages} {907} (\bibinfo {year} {2016})}\BibitemShut
  {NoStop}%
\bibitem [{\citenamefont {Mott}(1968)}]{MOTT1968_1}%
  \BibitemOpen
  \bibfield  {author} {\bibinfo {author} {\bibfnamefont {N.}~\bibnamefont
  {Mott}},\ }\href {\doibase http://dx.doi.org/10.1016/0022-3093(68)90002-1}
  {\bibfield  {journal} {\bibinfo  {journal} {Journal of Non-Crystalline
  Solids}\ }\textbf {\bibinfo {volume} {1}},\ \bibinfo {pages} {1 } (\bibinfo
  {year} {1968})}\BibitemShut {NoStop}%
\bibitem [{\citenamefont {Prelov{\v{s}}ek}\ \emph {et~al.}(2016)\citenamefont
  {Prelov{\v{s}}ek}, \citenamefont {Bari{\v{s}}i{\'{c}}},\ and\ \citenamefont
  {{\v{Z}}nidari{\v{c}}}}]{prelovsek216}%
  \BibitemOpen
  \bibfield  {author} {\bibinfo {author} {\bibfnamefont {P.}~\bibnamefont
  {Prelov{\v{s}}ek}}, \bibinfo {author} {\bibfnamefont {O.~S.}\ \bibnamefont
  {Bari{\v{s}}i{\'{c}}}}, \ and\ \bibinfo {author} {\bibfnamefont
  {M.}~\bibnamefont {{\v{Z}}nidari{\v{c}}}},\ }\href {\doibase
  10.1103/PhysRevB.94.241104} {\bibfield  {journal} {\bibinfo  {journal}
  {Physical Review B}\ }\textbf {\bibinfo {volume} {94}},\ \bibinfo {pages}
  {241104} (\bibinfo {year} {2016})}\BibitemShut {NoStop}%
\bibitem [{\citenamefont {Parameswaran}\ and\ \citenamefont
  {Gopalakrishnan}(2017)}]{parameswaran2017}%
  \BibitemOpen
  \bibfield  {author} {\bibinfo {author} {\bibfnamefont {S.~A.}\ \bibnamefont
  {Parameswaran}}\ and\ \bibinfo {author} {\bibfnamefont {S.}~\bibnamefont
  {Gopalakrishnan}},\ }\href {\doibase 10.1103/PhysRevB.95.024201} {\bibfield
  {journal} {\bibinfo  {journal} {Phys. Rev. B}\ }\textbf {\bibinfo {volume}
  {95}},\ \bibinfo {pages} {024201} (\bibinfo {year} {2017})}\BibitemShut
  {NoStop}%
\bibitem [{\citenamefont {Bon\ifmmode~\check{c}\else \v{c}\fi{}a}\ and\
  \citenamefont {Mierzejewski}(2017)}]{bonca2017}%
  \BibitemOpen
  \bibfield  {author} {\bibinfo {author} {\bibfnamefont {J.}~\bibnamefont
  {Bon\ifmmode~\check{c}\else \v{c}\fi{}a}}\ and\ \bibinfo {author}
  {\bibfnamefont {M.}~\bibnamefont {Mierzejewski}},\ }\href {\doibase
  10.1103/PhysRevB.95.214201} {\bibfield  {journal} {\bibinfo  {journal} {Phys.
  Rev. B}\ }\textbf {\bibinfo {volume} {95}},\ \bibinfo {pages} {214201}
  (\bibinfo {year} {2017})}\BibitemShut {NoStop}%
\bibitem [{\citenamefont {Serbyn}\ \emph
  {et~al.}(2013{\natexlab{b}})\citenamefont {Serbyn}, \citenamefont
  {Papi\'{c}},\ and\ \citenamefont {Abanin}}]{Serbyn13}%
  \BibitemOpen
  \bibfield  {author} {\bibinfo {author} {\bibfnamefont {M.}~\bibnamefont
  {Serbyn}}, \bibinfo {author} {\bibfnamefont {Z.}~\bibnamefont {Papi\'{c}}}, \
  and\ \bibinfo {author} {\bibfnamefont {D.~A.}\ \bibnamefont {Abanin}},\
  }\href {\doibase 10.1103/PhysRevLett.110.260601} {\bibfield  {journal}
  {\bibinfo  {journal} {Phys. Rev. Lett.}\ }\textbf {\bibinfo {volume} {110}},\
  \bibinfo {pages} {260601} (\bibinfo {year} {2013}{\natexlab{b}})}\BibitemShut
  {NoStop}%
\bibitem [{\citenamefont {{Imbrie}}\ \emph {et~al.}(2016)\citenamefont
  {{Imbrie}}, \citenamefont {{Ros}},\ and\ \citenamefont
  {{Scardicchio}}}]{imbrie2017}%
  \BibitemOpen
  \bibfield  {author} {\bibinfo {author} {\bibfnamefont {J.~Z.}\ \bibnamefont
  {{Imbrie}}}, \bibinfo {author} {\bibfnamefont {V.}~\bibnamefont {{Ros}}}, \
  and\ \bibinfo {author} {\bibfnamefont {A.}~\bibnamefont {{Scardicchio}}},\
  }\href@noop {} {\bibfield  {journal} {\bibinfo  {journal} {ArXiv e-prints}\ }
  (\bibinfo {year} {2016})},\ \Eprint {http://arxiv.org/abs/1609.08076}
  {arXiv:1609.08076 [cond-mat.dis-nn]} \BibitemShut {NoStop}%
\bibitem [{\citenamefont {Srednicki}(1994)}]{srednicki94}%
  \BibitemOpen
  \bibfield  {author} {\bibinfo {author} {\bibfnamefont {M.}~\bibnamefont
  {Srednicki}},\ }\href {\doibase 10.1103/PhysRevE.50.888} {\bibfield
  {journal} {\bibinfo  {journal} {Phys. Rev. E}\ }\textbf {\bibinfo {volume}
  {50}},\ \bibinfo {pages} {888} (\bibinfo {year} {1994})}\BibitemShut
  {NoStop}%
\bibitem [{\citenamefont {Abanin}\ \emph {et~al.}(2017)\citenamefont {Abanin},
  \citenamefont {De~Roeck}, \citenamefont {Ho},\ and\ \citenamefont
  {Huveneers}}]{abanin2017}%
  \BibitemOpen
  \bibfield  {author} {\bibinfo {author} {\bibfnamefont {D.~A.}\ \bibnamefont
  {Abanin}}, \bibinfo {author} {\bibfnamefont {W.}~\bibnamefont {De~Roeck}},
  \bibinfo {author} {\bibfnamefont {W.~W.}\ \bibnamefont {Ho}}, \ and\ \bibinfo
  {author} {\bibfnamefont {F.~m.~c.}\ \bibnamefont {Huveneers}},\ }\href
  {\doibase 10.1103/PhysRevB.95.014112} {\bibfield  {journal} {\bibinfo
  {journal} {Phys. Rev. B}\ }\textbf {\bibinfo {volume} {95}},\ \bibinfo
  {pages} {014112} (\bibinfo {year} {2017})}\BibitemShut {NoStop}%
\bibitem [{\citenamefont {Abanin}\ \emph {et~al.}(2015)\citenamefont {Abanin},
  \citenamefont {De~Roeck},\ and\ \citenamefont {Huveneers}}]{abanin2015}%
  \BibitemOpen
  \bibfield  {author} {\bibinfo {author} {\bibfnamefont {D.~A.}\ \bibnamefont
  {Abanin}}, \bibinfo {author} {\bibfnamefont {W.}~\bibnamefont {De~Roeck}}, \
  and\ \bibinfo {author} {\bibfnamefont {F.}~\bibnamefont {Huveneers}},\ }\href
  {\doibase 10.1103/PhysRevLett.115.256803} {\bibfield  {journal} {\bibinfo
  {journal} {Phys. Rev. Lett.}\ }\textbf {\bibinfo {volume} {115}},\ \bibinfo
  {pages} {256803} (\bibinfo {year} {2015})}\BibitemShut {NoStop}%
\bibitem [{\citenamefont {Lazarides}\ \emph {et~al.}(2015)\citenamefont
  {Lazarides}, \citenamefont {Das},\ and\ \citenamefont
  {Moessner}}]{Lazarides15}%
  \BibitemOpen
  \bibfield  {author} {\bibinfo {author} {\bibfnamefont {A.}~\bibnamefont
  {Lazarides}}, \bibinfo {author} {\bibfnamefont {A.}~\bibnamefont {Das}}, \
  and\ \bibinfo {author} {\bibfnamefont {R.}~\bibnamefont {Moessner}},\ }\href
  {\doibase 10.1103/PhysRevLett.115.030402} {\bibfield  {journal} {\bibinfo
  {journal} {Phys. Rev. Lett.}\ }\textbf {\bibinfo {volume} {115}},\ \bibinfo
  {pages} {030402} (\bibinfo {year} {2015})}\BibitemShut {NoStop}%
\bibitem [{\citenamefont {Ponte}\ \emph {et~al.}(2015)\citenamefont {Ponte},
  \citenamefont {Papi\ifmmode~\acute{c}\else \'{c}\fi{}}, \citenamefont
  {Huveneers},\ and\ \citenamefont {Abanin}}]{Ponte2015}%
  \BibitemOpen
  \bibfield  {author} {\bibinfo {author} {\bibfnamefont {P.}~\bibnamefont
  {Ponte}}, \bibinfo {author} {\bibfnamefont {Z.}~\bibnamefont
  {Papi\ifmmode~\acute{c}\else \'{c}\fi{}}}, \bibinfo {author} {\bibfnamefont
  {F.~m.~c.}\ \bibnamefont {Huveneers}}, \ and\ \bibinfo {author}
  {\bibfnamefont {D.~A.}\ \bibnamefont {Abanin}},\ }\href {\doibase
  10.1103/PhysRevLett.114.140401} {\bibfield  {journal} {\bibinfo  {journal}
  {Phys. Rev. Lett.}\ }\textbf {\bibinfo {volume} {114}},\ \bibinfo {pages}
  {140401} (\bibinfo {year} {2015})}\BibitemShut {NoStop}%
\bibitem [{\citenamefont {Kuwahara}\ \emph {et~al.}(2016)\citenamefont
  {Kuwahara}, \citenamefont {Mori},\ and\ \citenamefont
  {Saito}}]{Kuwahara2016}%
  \BibitemOpen
  \bibfield  {author} {\bibinfo {author} {\bibfnamefont {T.}~\bibnamefont
  {Kuwahara}}, \bibinfo {author} {\bibfnamefont {T.}~\bibnamefont {Mori}}, \
  and\ \bibinfo {author} {\bibfnamefont {K.}~\bibnamefont {Saito}},\ }\href
  {\doibase http://doi.org/10.1016/j.aop.2016.01.012} {\bibfield  {journal}
  {\bibinfo  {journal} {Annals of Physics}\ }\textbf {\bibinfo {volume}
  {367}},\ \bibinfo {pages} {96 } (\bibinfo {year} {2016})}\BibitemShut
  {NoStop}%
\bibitem [{\citenamefont {Bordia}\ \emph {et~al.}(2017)\citenamefont {Bordia},
  \citenamefont {Luschen}, \citenamefont {Schneider}, \citenamefont {Knap},\
  and\ \citenamefont {Bloch}}]{Bordia2017}%
  \BibitemOpen
  \bibfield  {author} {\bibinfo {author} {\bibfnamefont {P.}~\bibnamefont
  {Bordia}}, \bibinfo {author} {\bibfnamefont {H.}~\bibnamefont {Luschen}},
  \bibinfo {author} {\bibfnamefont {U.}~\bibnamefont {Schneider}}, \bibinfo
  {author} {\bibfnamefont {M.}~\bibnamefont {Knap}}, \ and\ \bibinfo {author}
  {\bibfnamefont {I.}~\bibnamefont {Bloch}},\ }\href
  {http://dx.doi.org/10.1038/nphys4020} {\bibfield  {journal} {\bibinfo
  {journal} {Nat Phys}\ }\textbf {\bibinfo {volume} {advance online
  publication}} (\bibinfo {year} {2017})},\ \bibinfo {note}
  {article}\BibitemShut {NoStop}%
\bibitem [{\citenamefont {Rehn}\ \emph {et~al.}(2016)\citenamefont {Rehn},
  \citenamefont {Lazarides}, \citenamefont {Pollmann},\ and\ \citenamefont
  {Moessner}}]{Rehn2016}%
  \BibitemOpen
  \bibfield  {author} {\bibinfo {author} {\bibfnamefont {J.}~\bibnamefont
  {Rehn}}, \bibinfo {author} {\bibfnamefont {A.}~\bibnamefont {Lazarides}},
  \bibinfo {author} {\bibfnamefont {F.}~\bibnamefont {Pollmann}}, \ and\
  \bibinfo {author} {\bibfnamefont {R.}~\bibnamefont {Moessner}},\ }\href
  {\doibase 10.1103/PhysRevB.94.020201} {\bibfield  {journal} {\bibinfo
  {journal} {Phys. Rev. B}\ }\textbf {\bibinfo {volume} {94}},\ \bibinfo
  {pages} {020201} (\bibinfo {year} {2016})}\BibitemShut {NoStop}%
\bibitem [{\citenamefont {Choi}\ \emph {et~al.}(2017)\citenamefont {Choi},
  \citenamefont {Choi}, \citenamefont {Landig}, \citenamefont {Kucsko},
  \citenamefont {Zhou}, \citenamefont {Isoya}, \citenamefont {Jelezko},
  \citenamefont {Onoda}, \citenamefont {Sumiya}, \citenamefont {Khemani},
  \citenamefont {von Keyserlingk}, \citenamefont {Yao}, \citenamefont
  {Demler},\ and\ \citenamefont {Lukin}}]{Choi2017}%
  \BibitemOpen
  \bibfield  {author} {\bibinfo {author} {\bibfnamefont {S.}~\bibnamefont
  {Choi}}, \bibinfo {author} {\bibfnamefont {J.}~\bibnamefont {Choi}}, \bibinfo
  {author} {\bibfnamefont {R.}~\bibnamefont {Landig}}, \bibinfo {author}
  {\bibfnamefont {G.}~\bibnamefont {Kucsko}}, \bibinfo {author} {\bibfnamefont
  {H.}~\bibnamefont {Zhou}}, \bibinfo {author} {\bibfnamefont {J.}~\bibnamefont
  {Isoya}}, \bibinfo {author} {\bibfnamefont {F.}~\bibnamefont {Jelezko}},
  \bibinfo {author} {\bibfnamefont {S.}~\bibnamefont {Onoda}}, \bibinfo
  {author} {\bibfnamefont {H.}~\bibnamefont {Sumiya}}, \bibinfo {author}
  {\bibfnamefont {V.}~\bibnamefont {Khemani}}, \bibinfo {author} {\bibfnamefont
  {C.}~\bibnamefont {von Keyserlingk}}, \bibinfo {author} {\bibfnamefont
  {N.~Y.}\ \bibnamefont {Yao}}, \bibinfo {author} {\bibfnamefont
  {E.}~\bibnamefont {Demler}}, \ and\ \bibinfo {author} {\bibfnamefont {M.~D.}\
  \bibnamefont {Lukin}},\ }\href {http://dx.doi.org/10.1038/nature21426}
  {\bibfield  {journal} {\bibinfo  {journal} {Nature}\ }\textbf {\bibinfo
  {volume} {543}},\ \bibinfo {pages} {221} (\bibinfo {year} {2017})},\ \bibinfo
  {note} {letter}\BibitemShut {NoStop}%
\bibitem [{\citenamefont {Zhang}\ \emph {et~al.}(2017)\citenamefont {Zhang},
  \citenamefont {Hess}, \citenamefont {Kyprianidis}, \citenamefont {Becker},
  \citenamefont {Lee}, \citenamefont {Smith}, \citenamefont {Pagano},
  \citenamefont {Potirniche}, \citenamefont {Potter}, \citenamefont
  {Vishwanath}, \citenamefont {Yao},\ and\ \citenamefont {Monroe}}]{Zhang2017}%
  \BibitemOpen
  \bibfield  {author} {\bibinfo {author} {\bibfnamefont {J.}~\bibnamefont
  {Zhang}}, \bibinfo {author} {\bibfnamefont {P.~W.}\ \bibnamefont {Hess}},
  \bibinfo {author} {\bibfnamefont {A.}~\bibnamefont {Kyprianidis}}, \bibinfo
  {author} {\bibfnamefont {P.}~\bibnamefont {Becker}}, \bibinfo {author}
  {\bibfnamefont {A.}~\bibnamefont {Lee}}, \bibinfo {author} {\bibfnamefont
  {J.}~\bibnamefont {Smith}}, \bibinfo {author} {\bibfnamefont
  {G.}~\bibnamefont {Pagano}}, \bibinfo {author} {\bibfnamefont {I.-D.}\
  \bibnamefont {Potirniche}}, \bibinfo {author} {\bibfnamefont {A.~C.}\
  \bibnamefont {Potter}}, \bibinfo {author} {\bibfnamefont {A.}~\bibnamefont
  {Vishwanath}}, \bibinfo {author} {\bibfnamefont {N.~Y.}\ \bibnamefont {Yao}},
  \ and\ \bibinfo {author} {\bibfnamefont {C.}~\bibnamefont {Monroe}},\ }\href
  {http://dx.doi.org/10.1038/nature21413} {\bibfield  {journal} {\bibinfo
  {journal} {Nature}\ }\textbf {\bibinfo {volume} {543}},\ \bibinfo {pages}
  {217} (\bibinfo {year} {2017})},\ \bibinfo {note} {letter}\BibitemShut
  {NoStop}%
\bibitem [{\citenamefont {Sacha}(2015{\natexlab{a}})}]{Sacha2015}%
  \BibitemOpen
  \bibfield  {author} {\bibinfo {author} {\bibfnamefont {K.}~\bibnamefont
  {Sacha}},\ }\href {\doibase 10.1103/PhysRevA.91.033617} {\bibfield  {journal}
  {\bibinfo  {journal} {Phys. Rev. A}\ }\textbf {\bibinfo {volume} {91}},\
  \bibinfo {pages} {033617} (\bibinfo {year} {2015}{\natexlab{a}})}\BibitemShut
  {NoStop}%
\bibitem [{\citenamefont {Khemani}\ \emph {et~al.}(2016)\citenamefont
  {Khemani}, \citenamefont {Lazarides}, \citenamefont {Moessner},\ and\
  \citenamefont {Sondhi}}]{Khemani16}%
  \BibitemOpen
  \bibfield  {author} {\bibinfo {author} {\bibfnamefont {V.}~\bibnamefont
  {Khemani}}, \bibinfo {author} {\bibfnamefont {A.}~\bibnamefont {Lazarides}},
  \bibinfo {author} {\bibfnamefont {R.}~\bibnamefont {Moessner}}, \ and\
  \bibinfo {author} {\bibfnamefont {S.~L.}\ \bibnamefont {Sondhi}},\ }\href
  {\doibase 10.1103/PhysRevLett.116.250401} {\bibfield  {journal} {\bibinfo
  {journal} {Phys. Rev. Lett.}\ }\textbf {\bibinfo {volume} {116}},\ \bibinfo
  {pages} {250401} (\bibinfo {year} {2016})}\BibitemShut {NoStop}%
\bibitem [{\citenamefont {von Keyserlingk}\ \emph {et~al.}(2016)\citenamefont
  {von Keyserlingk}, \citenamefont {Khemani},\ and\ \citenamefont
  {Sondhi}}]{Keyserlingk16}%
  \BibitemOpen
  \bibfield  {author} {\bibinfo {author} {\bibfnamefont {C.~W.}\ \bibnamefont
  {von Keyserlingk}}, \bibinfo {author} {\bibfnamefont {V.}~\bibnamefont
  {Khemani}}, \ and\ \bibinfo {author} {\bibfnamefont {S.~L.}\ \bibnamefont
  {Sondhi}},\ }\href {\doibase 10.1103/PhysRevB.94.085112} {\bibfield
  {journal} {\bibinfo  {journal} {Phys. Rev. B}\ }\textbf {\bibinfo {volume}
  {94}},\ \bibinfo {pages} {085112} (\bibinfo {year} {2016})}\BibitemShut
  {NoStop}%
\bibitem [{\citenamefont {Else}\ \emph {et~al.}(2016)\citenamefont {Else},
  \citenamefont {Bauer},\ and\ \citenamefont {Nayak}}]{ElseFTC}%
  \BibitemOpen
  \bibfield  {author} {\bibinfo {author} {\bibfnamefont {D.~V.}\ \bibnamefont
  {Else}}, \bibinfo {author} {\bibfnamefont {B.}~\bibnamefont {Bauer}}, \ and\
  \bibinfo {author} {\bibfnamefont {C.}~\bibnamefont {Nayak}},\ }\href
  {\doibase 10.1103/PhysRevLett.117.090402} {\bibfield  {journal} {\bibinfo
  {journal} {Phys. Rev. Lett.}\ }\textbf {\bibinfo {volume} {117}},\ \bibinfo
  {pages} {090402} (\bibinfo {year} {2016})}\BibitemShut {NoStop}%
\bibitem [{\citenamefont {Yao}\ \emph {et~al.}(2017)\citenamefont {Yao},
  \citenamefont {Potter}, \citenamefont {Potirniche},\ and\ \citenamefont
  {Vishwanath}}]{Yao2017}%
  \BibitemOpen
  \bibfield  {author} {\bibinfo {author} {\bibfnamefont {N.~Y.}\ \bibnamefont
  {Yao}}, \bibinfo {author} {\bibfnamefont {A.~C.}\ \bibnamefont {Potter}},
  \bibinfo {author} {\bibfnamefont {I.-D.}\ \bibnamefont {Potirniche}}, \ and\
  \bibinfo {author} {\bibfnamefont {A.}~\bibnamefont {Vishwanath}},\ }\href
  {\doibase 10.1103/PhysRevLett.118.030401} {\bibfield  {journal} {\bibinfo
  {journal} {Phys. Rev. Lett.}\ }\textbf {\bibinfo {volume} {118}},\ \bibinfo
  {pages} {030401} (\bibinfo {year} {2017})}\BibitemShut {NoStop}%
\bibitem [{\citenamefont {Nayak}(2017)}]{Nayak2017}%
  \BibitemOpen
  \bibfield  {author} {\bibinfo {author} {\bibfnamefont {C.}~\bibnamefont
  {Nayak}},\ }\href {http://dx.doi.org/10.1038/543185a} {\bibfield  {journal}
  {\bibinfo  {journal} {Nature}\ }\textbf {\bibinfo {volume} {543}},\ \bibinfo
  {pages} {185} (\bibinfo {year} {2017})},\ \bibinfo {note} {news \&
  Views}\BibitemShut {NoStop}%
\bibitem [{\citenamefont {Else}\ \emph {et~al.}(2017)\citenamefont {Else},
  \citenamefont {Bauer},\ and\ \citenamefont {Nayak}}]{Else17prx}%
  \BibitemOpen
  \bibfield  {author} {\bibinfo {author} {\bibfnamefont {D.~V.}\ \bibnamefont
  {Else}}, \bibinfo {author} {\bibfnamefont {B.}~\bibnamefont {Bauer}}, \ and\
  \bibinfo {author} {\bibfnamefont {C.}~\bibnamefont {Nayak}},\ }\href
  {\doibase 10.1103/PhysRevX.7.011026} {\bibfield  {journal} {\bibinfo
  {journal} {Phys. Rev. X}\ }\textbf {\bibinfo {volume} {7}},\ \bibinfo {pages}
  {011026} (\bibinfo {year} {2017})}\BibitemShut {NoStop}%
\bibitem [{\citenamefont {{Huang}}\ \emph {et~al.}(2017)\citenamefont
  {{Huang}}, \citenamefont {{Wu}},\ and\ \citenamefont {{Liu}}}]{Huang2017}%
  \BibitemOpen
  \bibfield  {author} {\bibinfo {author} {\bibfnamefont {B.}~\bibnamefont
  {{Huang}}}, \bibinfo {author} {\bibfnamefont {Y.-H.}\ \bibnamefont {{Wu}}}, \
  and\ \bibinfo {author} {\bibfnamefont {W.~V.}\ \bibnamefont {{Liu}}},\
  }\href@noop {} {\bibfield  {journal} {\bibinfo  {journal} {ArXiv e-prints}\ }
  (\bibinfo {year} {2017})},\ \Eprint {http://arxiv.org/abs/1703.04663}
  {arXiv:1703.04663 [cond-mat.quant-gas]} \BibitemShut {NoStop}%
\bibitem [{\citenamefont {Sacha}(2015{\natexlab{b}})}]{Sacha15a}%
  \BibitemOpen
  \bibfield  {author} {\bibinfo {author} {\bibfnamefont {K.}~\bibnamefont
  {Sacha}},\ }\href {\doibase 10.1038/srep10787} {\bibfield  {journal}
  {\bibinfo  {journal} {Scientific Reports}\ }\textbf {\bibinfo {volume} {5}},\
  \bibinfo {pages} {10787} (\bibinfo {year} {2015}{\natexlab{b}})}\BibitemShut
  {NoStop}%
\bibitem [{\citenamefont {Guo}\ \emph {et~al.}(2013)\citenamefont {Guo},
  \citenamefont {Marthaler},\ and\ \citenamefont {Sch\"on}}]{Guo2013}%
  \BibitemOpen
  \bibfield  {author} {\bibinfo {author} {\bibfnamefont {L.}~\bibnamefont
  {Guo}}, \bibinfo {author} {\bibfnamefont {M.}~\bibnamefont {Marthaler}}, \
  and\ \bibinfo {author} {\bibfnamefont {G.}~\bibnamefont {Sch\"on}},\ }\href
  {\doibase 10.1103/PhysRevLett.111.205303} {\bibfield  {journal} {\bibinfo
  {journal} {Phys. Rev. Lett.}\ }\textbf {\bibinfo {volume} {111}},\ \bibinfo
  {pages} {205303} (\bibinfo {year} {2013})}\BibitemShut {NoStop}%
\bibitem [{\citenamefont {Steane}\ \emph {et~al.}(1995)\citenamefont {Steane},
  \citenamefont {Szriftgiser}, \citenamefont {Desbiolles},\ and\ \citenamefont
  {Dalibard}}]{Steane95}%
  \BibitemOpen
  \bibfield  {author} {\bibinfo {author} {\bibfnamefont {A.}~\bibnamefont
  {Steane}}, \bibinfo {author} {\bibfnamefont {P.}~\bibnamefont {Szriftgiser}},
  \bibinfo {author} {\bibfnamefont {P.}~\bibnamefont {Desbiolles}}, \ and\
  \bibinfo {author} {\bibfnamefont {J.}~\bibnamefont {Dalibard}},\ }\href
  {\doibase 10.1103/PhysRevLett.74.4972} {\bibfield  {journal} {\bibinfo
  {journal} {Phys. Rev. Lett.}\ }\textbf {\bibinfo {volume} {74}},\ \bibinfo
  {pages} {4972} (\bibinfo {year} {1995})}\BibitemShut {NoStop}%
\bibitem [{\citenamefont {Buchleitner}\ \emph {et~al.}(2002)\citenamefont
  {Buchleitner}, \citenamefont {Delande},\ and\ \citenamefont
  {Zakrzewski}}]{Buchleitner2002}%
  \BibitemOpen
  \bibfield  {author} {\bibinfo {author} {\bibfnamefont {A.}~\bibnamefont
  {Buchleitner}}, \bibinfo {author} {\bibfnamefont {D.}~\bibnamefont
  {Delande}}, \ and\ \bibinfo {author} {\bibfnamefont {J.}~\bibnamefont
  {Zakrzewski}},\ }\href
  {http://www.sciencedirect.com/science/article/pii/S0370157302002703}
  {\bibfield  {journal} {\bibinfo  {journal} {Physics reports}\ }\textbf
  {\bibinfo {volume} {368}},\ \bibinfo {pages} {409} (\bibinfo {year}
  {2002})}\BibitemShut {NoStop}%
\bibitem [{\citenamefont {Clifford}\ \emph {et~al.}(1998)\citenamefont
  {Clifford}, \citenamefont {Arlt}, \citenamefont {Courtial},\ and\
  \citenamefont {Dholakia}}]{Clifford1998}%
  \BibitemOpen
  \bibfield  {author} {\bibinfo {author} {\bibfnamefont {M.}~\bibnamefont
  {Clifford}}, \bibinfo {author} {\bibfnamefont {J.}~\bibnamefont {Arlt}},
  \bibinfo {author} {\bibfnamefont {J.}~\bibnamefont {Courtial}}, \ and\
  \bibinfo {author} {\bibfnamefont {K.}~\bibnamefont {Dholakia}},\ }\href
  {\doibase http://doi.org/10.1016/S0030-4018(98)00464-7} {\bibfield  {journal}
  {\bibinfo  {journal} {Optics Communications}\ }\textbf {\bibinfo {volume}
  {156}},\ \bibinfo {pages} {300 } (\bibinfo {year} {1998})}\BibitemShut
  {NoStop}%
\bibitem [{\citenamefont {Wright}\ \emph {et~al.}(2000)\citenamefont {Wright},
  \citenamefont {Arlt},\ and\ \citenamefont {Dholakia}}]{Wright2000}%
  \BibitemOpen
  \bibfield  {author} {\bibinfo {author} {\bibfnamefont {E.~M.}\ \bibnamefont
  {Wright}}, \bibinfo {author} {\bibfnamefont {J.}~\bibnamefont {Arlt}}, \ and\
  \bibinfo {author} {\bibfnamefont {K.}~\bibnamefont {Dholakia}},\ }\href
  {\doibase 10.1103/PhysRevA.63.013608} {\bibfield  {journal} {\bibinfo
  {journal} {Phys. Rev. A}\ }\textbf {\bibinfo {volume} {63}},\ \bibinfo
  {pages} {013608} (\bibinfo {year} {2000})}\BibitemShut {NoStop}%
\bibitem [{\citenamefont {Wang}\ \emph {et~al.}(2009)\citenamefont {Wang},
  \citenamefont {Zhang},\ and\ \citenamefont {Lin}}]{Wang2009}%
  \BibitemOpen
  \bibfield  {author} {\bibinfo {author} {\bibfnamefont {Z.~Y.}\ \bibnamefont
  {Wang}}, \bibinfo {author} {\bibfnamefont {Z.}~\bibnamefont {Zhang}}, \ and\
  \bibinfo {author} {\bibfnamefont {Q.}~\bibnamefont {Lin}},\ }\href {\doibase
  10.1140/epjd/e2009-00116-7} {\bibfield  {journal} {\bibinfo  {journal} {The
  European Physical Journal D}\ }\textbf {\bibinfo {volume} {53}},\ \bibinfo
  {pages} {127} (\bibinfo {year} {2009})}\BibitemShut {NoStop}%
\bibitem [{\citenamefont {Lichtenberg}\ and\ \citenamefont
  {Lieberman}(1992)}]{Lichtenberg1992}%
  \BibitemOpen
  \bibfield  {author} {\bibinfo {author} {\bibfnamefont {A.}~\bibnamefont
  {Lichtenberg}}\ and\ \bibinfo {author} {\bibfnamefont {M.}~\bibnamefont
  {Lieberman}},\ }\href {https://books.google.pl/books?id=2ssPAQAAMAAJ} {\emph
  {\bibinfo {title} {Regular and chaotic dynamics}}},\ Applied mathematical
  sciences\ (\bibinfo  {publisher} {Springer-Verlag},\ \bibinfo {year}
  {1992})\BibitemShut {NoStop}%
\bibitem [{\citenamefont {Berman}\ and\ \citenamefont
  {Zaslavsky}(1977)}]{Berman1977}%
  \BibitemOpen
  \bibfield  {author} {\bibinfo {author} {\bibfnamefont {G.}~\bibnamefont
  {Berman}}\ and\ \bibinfo {author} {\bibfnamefont {G.}~\bibnamefont
  {Zaslavsky}},\ }\href {\doibase 10.1016/0375-9601(77)90618-1} {\bibfield
  {journal} {\bibinfo  {journal} {Physics Letters A}\ }\textbf {\bibinfo
  {volume} {61}},\ \bibinfo {pages} {295} (\bibinfo {year} {1977})}\BibitemShut
  {NoStop}%
\bibitem [{\citenamefont {Sacha}\ and\ \citenamefont
  {Delande}(2016)}]{sacha16}%
  \BibitemOpen
  \bibfield  {author} {\bibinfo {author} {\bibfnamefont {K.}~\bibnamefont
  {Sacha}}\ and\ \bibinfo {author} {\bibfnamefont {D.}~\bibnamefont
  {Delande}},\ }\href {\doibase 10.1103/PhysRevA.94.023633} {\bibfield
  {journal} {\bibinfo  {journal} {Phys. Rev. A}\ }\textbf {\bibinfo {volume}
  {94}},\ \bibinfo {pages} {023633} (\bibinfo {year} {2016})}\BibitemShut
  {NoStop}%
\bibitem [{\citenamefont {Guo}\ \emph {et~al.}(2016)\citenamefont {Guo},
  \citenamefont {Liu},\ and\ \citenamefont {Marthaler}}]{Guo2016a}%
  \BibitemOpen
  \bibfield  {author} {\bibinfo {author} {\bibfnamefont {L.}~\bibnamefont
  {Guo}}, \bibinfo {author} {\bibfnamefont {M.}~\bibnamefont {Liu}}, \ and\
  \bibinfo {author} {\bibfnamefont {M.}~\bibnamefont {Marthaler}},\ }\href
  {\doibase 10.1103/PhysRevA.93.053616} {\bibfield  {journal} {\bibinfo
  {journal} {Phys. Rev. A}\ }\textbf {\bibinfo {volume} {93}},\ \bibinfo
  {pages} {053616} (\bibinfo {year} {2016})}\BibitemShut {NoStop}%
\bibitem [{\citenamefont {Szriftgiser}\ \emph {et~al.}(1996)\citenamefont
  {Szriftgiser}, \citenamefont {Gu\'ery-Odelin}, \citenamefont {Arndt},\ and\
  \citenamefont {Dalibard}}]{Szriftgiser:Mirror:PRL96}%
  \BibitemOpen
  \bibfield  {author} {\bibinfo {author} {\bibfnamefont {P.}~\bibnamefont
  {Szriftgiser}}, \bibinfo {author} {\bibfnamefont {D.}~\bibnamefont
  {Gu\'ery-Odelin}}, \bibinfo {author} {\bibfnamefont {M.}~\bibnamefont
  {Arndt}}, \ and\ \bibinfo {author} {\bibfnamefont {J.}~\bibnamefont
  {Dalibard}},\ }\href {\doibase 10.1103/PhysRevLett.77.4} {\bibfield
  {journal} {\bibinfo  {journal} {Phys. Rev. Lett.}\ }\textbf {\bibinfo
  {volume} {77}},\ \bibinfo {pages} {4} (\bibinfo {year} {1996})}\BibitemShut
  {NoStop}%
\bibitem [{\citenamefont {Yao}\ \emph {et~al.}(2014)\citenamefont {Yao},
  \citenamefont {Laumann}, \citenamefont {Gopalakrishnan}, \citenamefont
  {Knap}, \citenamefont {M\"uller}, \citenamefont {Demler},\ and\ \citenamefont
  {Lukin}}]{yao2014}%
  \BibitemOpen
  \bibfield  {author} {\bibinfo {author} {\bibfnamefont {N.~Y.}\ \bibnamefont
  {Yao}}, \bibinfo {author} {\bibfnamefont {C.~R.}\ \bibnamefont {Laumann}},
  \bibinfo {author} {\bibfnamefont {S.}~\bibnamefont {Gopalakrishnan}},
  \bibinfo {author} {\bibfnamefont {M.}~\bibnamefont {Knap}}, \bibinfo {author}
  {\bibfnamefont {M.}~\bibnamefont {M\"uller}}, \bibinfo {author}
  {\bibfnamefont {E.~A.}\ \bibnamefont {Demler}}, \ and\ \bibinfo {author}
  {\bibfnamefont {M.~D.}\ \bibnamefont {Lukin}},\ }\href {\doibase
  10.1103/PhysRevLett.113.243002} {\bibfield  {journal} {\bibinfo  {journal}
  {Phys. Rev. Lett.}\ }\textbf {\bibinfo {volume} {113}},\ \bibinfo {pages}
  {243002} (\bibinfo {year} {2014})}\BibitemShut {NoStop}%
\bibitem [{\citenamefont {{Nandkishore}}\ and\ \citenamefont
  {{Sondhi}}(2017)}]{nand2017}%
  \BibitemOpen
  \bibfield  {author} {\bibinfo {author} {\bibfnamefont {R.~M.}\ \bibnamefont
  {{Nandkishore}}}\ and\ \bibinfo {author} {\bibfnamefont {S.~L.}\ \bibnamefont
  {{Sondhi}}},\ }\href@noop {} {\bibfield  {journal} {\bibinfo  {journal}
  {ArXiv e-prints}\ } (\bibinfo {year} {2017})},\ \Eprint
  {http://arxiv.org/abs/1705.06290} {arXiv:1705.06290 [cond-mat.str-el]}
  \BibitemShut {NoStop}%
\bibitem [{\citenamefont {Luitz}\ \emph
  {et~al.}(2015{\natexlab{b}})\citenamefont {Luitz}, \citenamefont
  {Laflorencie},\ and\ \citenamefont {Alet}}]{luitz15}%
  \BibitemOpen
  \bibfield  {author} {\bibinfo {author} {\bibfnamefont {D.~J.}\ \bibnamefont
  {Luitz}}, \bibinfo {author} {\bibfnamefont {N.}~\bibnamefont {Laflorencie}},
  \ and\ \bibinfo {author} {\bibfnamefont {F.}~\bibnamefont {Alet}},\ }\href
  {\doibase 10.1103/PhysRevB.91.081103} {\bibfield  {journal} {\bibinfo
  {journal} {Phys. Rev. B}\ }\textbf {\bibinfo {volume} {91}},\ \bibinfo
  {pages} {081103} (\bibinfo {year} {2015}{\natexlab{b}})}\BibitemShut
  {NoStop}%
\bibitem [{\citenamefont {{Sacha}}\ and\ \citenamefont
  {{Zakrzewski}}(2017)}]{Sacha2017review}%
  \BibitemOpen
  \bibfield  {author} {\bibinfo {author} {\bibfnamefont {K.}~\bibnamefont
  {{Sacha}}}\ and\ \bibinfo {author} {\bibfnamefont {J.}~\bibnamefont
  {{Zakrzewski}}},\ }\href@noop {} {\bibfield  {journal} {\bibinfo  {journal}
  {ArXiv e-prints}\ } (\bibinfo {year} {2017})},\ \Eprint
  {http://arxiv.org/abs/1704.03735} {arXiv:1704.03735 [quant-ph]} \BibitemShut
  {NoStop}%
\end{thebibliography}
%

\end{document}